\documentclass[12pt]{article} 
\usepackage[utf8]{inputenc} 
\usepackage{lmodern}  
\usepackage{amsmath} 
\usepackage{amsfonts}  
\usepackage{amssymb}  
\usepackage[english]{babel} 
\usepackage{graphicx}
\usepackage{subfig}
\usepackage{authblk}

\usepackage[pdfauthor=L Glaser,pdftitle=Computer simulations of causal sets]{hyperref}

\usepackage[sorting=none,maxnames=5,firstinits=true,doi=true,url=true]{biblatex}
\newcommand{\md}{\mathrm{d}}
\newcommand{\mc}[1]{\mathcal{#1}}
\newcommand{\bb}{\bar\beta}

\newcommand{\G}[1]{\Gamma\left( #1 \right)}
\newcommand{\av}[1]{\left\langle #1 \right\rangle}
\newcommand{\Sa}{\mathcal{S}}

\title{Computer simulations of causal sets}
\author{L Glaser }
\affil{University of Vienna,\\ Boltzmanngasse 5, 1090 Vienna, Austria}

\begin{document}
\maketitle
\begin{abstract}
  This review introduces Markov Chain Monte Carlo simulations in causal set theory, with a focus on the study of the $2$d orders.
  It will first introduce the Benincasa-Dowker action on causal sets, and cover some musings on the philosophy of computer simulations.
  And then proceed to review results from the study of the $2$d orders, first their general phase transition and scaling behavior and then results on defining a wave function of the universe using these orders and on coupling the $2$d orders to an Ising like model.
  Including matter in this type shows a strong coupling between matter and geometry, that leads to new phase transitions.
  However, while the matter does induce a new phase transition, it does not change the order of the phase transition of geometry.
\end{abstract}

\tableofcontents

\section{Introduction} \label{sec:intro}
A causal set is spacetime reduced to a discrete causal structure.
To develop a theory of quantum gravity based on such a discretization requires us to quantize the dynamics of the theory.
The natural language for this is a path integral, that sums over causal sets.
To non-perturbatively study this path sum we can use computer simulations.

To study the path integral, which is fundamentally a quantum theory of Lorentzian spacetime, using computer simulations, it is turned into a statistical sum over Lorentzian spacetimes.
This is necessary since the causal set equivalent of the Einstein Hilbert action is purely real, and would thus lead to a complex weight factor, which is difficult to sample numerically.

Aside from the numerical methods focussed at in this review there are other paths to understand how causal sets can make up our universe.
For a full picture of the quantum theory we would like a covariant dynamics motivated intrinsically through the causal set, this approach is described more in the chapter `Covariant Causal Set Dynamics' written by Stav Zalel.
Another approach towards defining the quantum dynamics are sequential growth models, which constructs causal sets element by element, and thus define a measure on the space of all causal set, as introduced by Rideout and Sorkin in~\cite{Rideout:1999ub}.

This chapter will focus on computer simulations of a model system of causal sets, the $2$d orders. 
In the next section~\ref{sec:eve} we will discuss the interplay or entropy and action on the space of all partial orders, which is dominated by the Kleitman Rothschild orders, described in subsection~\ref{subsec:KR} and then introduce the $2$d orders as a restricted class of partial orders which are of special interest in causal set theory, in subsection~\ref{subsec:2dO}. This section closes with a short review of how the action on causal sets is derived, subsection~\ref{subsec:action}, and in subsection~\ref{subsec:MCMC} introduce Markov Chain Monte Carlo Simulations, the main technical tool used in simulating causal set theory.
The next section~\ref{sec:sims} introduces computer simulations on the $2$d orders, and splits into three subsections, each focussing on a different aspect.
The first subsection~\ref{subsec:2d} describes the simulations of pure $2$d orders, the study of their phase transitions and scalings. 
The next subsection~\ref{subsec:HH} describes work on computer simulations of a Hawking-Hartle wave function in the $2$d orders, and the last subsection~\ref{subsec:Ising} describes results on the $2$d orders coupled to the Ising model.
This chapter closes with a short outlook~\ref{sec:pers} on other open questions in simulations of the $2$d orders, and possible extensions of the model.

\section{Entropy vs Energy, the struggle of the path integral}\label{sec:eve}
Exploring the path integral over a theory of quantum gravity in computer simulations, means studying the interplay between the entropy in the space of configurations, and the action assigned to these configurations.
Changing the class of configurations can lead to a momentous change in the path integral, as impressively demonstrated by causal dynamical triangulations (CDT)~\cite{Ambjorn_Goerlich_Jurkiewicz_Loll_2012}.
While the input of CDT and dynamical triangulations (DT) is almost identical, the restriction of the path integral in CDT to only those configurations which allow for a definite time direction at each point, by imposing a preferred time foliation, changes the path integral completely.
While DT does not show a phase transition of higher order, and thus does not allow for a continuum limit, CDT does not only have such a phase transition, but also has a phase adjacent to this transition that shows $4$d continuum like behavior.

It is thus clear that explorations of the path integral over causal sets will need to consider both which class of causal sets to sum over, and with which action to weight them with.
This entropy of configurations is baked into the configuration space, it is the entropy on the space of all causal sets, so the counting of partial orders.

\subsection{Kleitman Rothschild orders} \label{subsec:KR}
\begin{figure}
  \includegraphics[width=\textwidth]{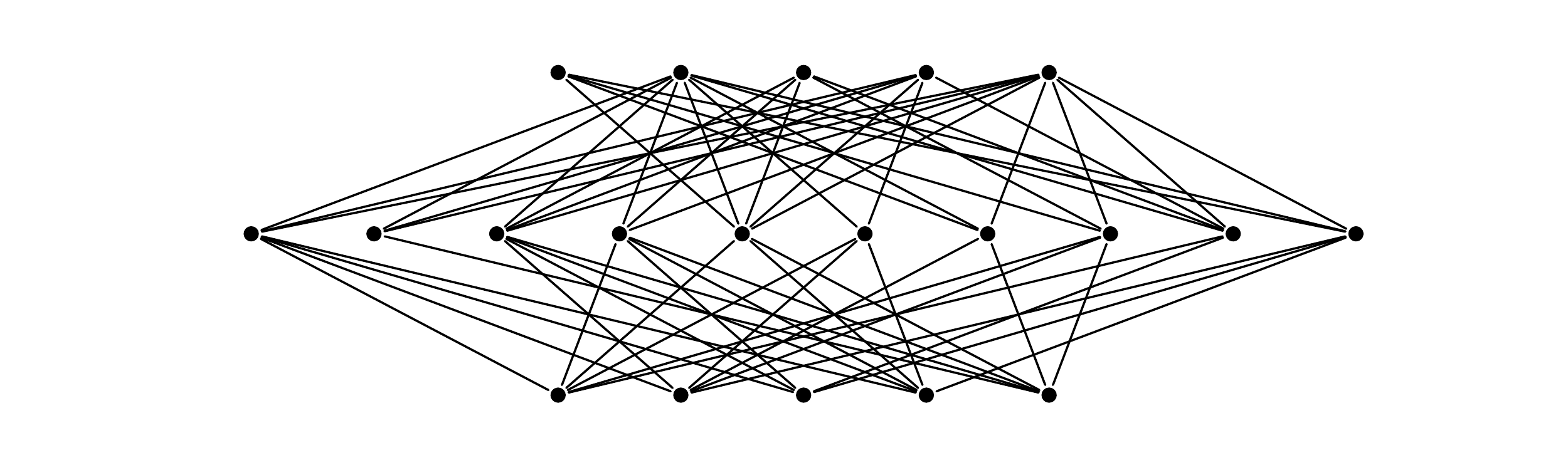}
  \caption{Example of a $20$ element Kleitman Rothschild order. }
\label{fig:KRorder}
\end{figure}

Mathematically speaking, a causal set is a partial order, and in the mathematical literature on partial orders, there is an impressive result by Kleitman and Rothschild, concerning the number of possible partial orders.
In~\cite{Kleitman_Rothschild_1975} they show that, in the large $N$ limit, the number of possible partial orders goes like $N^2/4$.
They prove this by establishing that this is the number of a particular type of three layer order, which dominates the number of all partial orders.
These orders, often called Kleitman Rothschild (KR) orders, are defined as follows; they consist of three layers, $L_1,L_2,L_3$, with $N/4+ \mathcal{O}(N^{1/2} \log{N})$ elements in layers $L_1,L_3$ and $N/2 + \mathcal{O}(N^{1/2}  \log{N})$ elements in layer $L_2$.
Each element in layer $L_2$ is to the future of half of the elements of $L_1$ and the past of half of the elements of $L_3$, and thus all elements of $L_1$ are in the past of all elements of $L_3$.
An example of a $20$ element KR order is shown in Figure~\ref{fig:KRorder}.

This mathematical result concerns the large $N$ limit, which leaves the question of how large $N$ can be before the KR orders dominate.
To explore this, Henson et al.~\cite{Henson_Rideout_Sorkin_Surya_2015} used Monte Carlo simulations to sample the space of all causal sets.
They found that KR orders are already $> 90\%$ of the orders sampled for $N=85$, which indicates that the large $N$ regime is already reached.
Since our universe consists of, assuming each element to be of Planck volume, roughly $10^{240}$ elements, the most likely partial orders of the size of the universe are clearly KR orders.

The entropy on the space of all causal sets is thus not in favor of manifoldlike causal sets, requiring a large energy to suppress these in the statistical path integral, as studied in computer simulations.
There is however hope that the case is different for the complex path integral.
It is possible to show that the integral over KR orders is suppressed by actions for which the number of links is the leading term~\cite{Mathur:2020hxl}, as is the case for most casual set actions, as introduced below.

\subsection{$2$d orders as causal sets}\label{subsec:2dO}
The class of all partial orders is very large, and as we discussed above dominated by the KR orders.
This motivated the search for a useful subclass of partial orders to study using numerics.
This class was found in \cite{Brightwell_Henson_Surya_2008} to be the 2d orders, defined in \cite{Winkler_1985}. 
A $N$ element $2$d order starts with the set $S$ or integers from $1$ to $N$, and two sets of labels $U=(u_1,u_2, \dots u_N)$, $V=(v_1,_2, \dots v_N)$ with $u_i,v_i \in S$ and $u_i\neq u_j, v_i \neq v_j$ unless $i=j$.
The sets $U,V$ are then total orders, with their ordering induced from the ordering on $S$, and we define a $N$ element $2$d order as the intersection $C= U \cap V$.
So $e_i \prec e_j$ iff $u_i<u_j$ and $v_i<v_j$, a useful example of such a $2$d order is a set of events in $2$d Minkowski spacetime, for which no $u_i$ or $v_i$ coincide.

The $2$d orders have several advantages for numerical studies of partial orders.
One is that, while they include a large variety of non-manifoldlike partial orders, it can be shown that the `most likely' $2$d orders all faithfully embed into $2$d Minkowski space~\cite{Winkler_1985,Brightwell_Henson_Surya_2008}.
These faithfully embedding orders are often also referred to as random orders.
They are thus a model that allows us to study spacetime of a fixed dimension through causal set theory.
In addition, the labelling by two numbers defines an embedding into $2$d Minkowski space, and thus allows us to plot the orders we study, and provides an additional tool to understand the system.

\subsection{Energy -- An action for a causal set}\label{subsec:action}
Since we have covered the entropy part of the story, it is now time to think about the energy of a causal set.
To take a path sum over causal sets, we need to define a causal set action, that weights the individual causal sets.
In principle, we have complete freedom to define the action, however to recover general relativity in the limit of manifold-like causal sets, it seems prudent to demand that the action should approximate the Einstein Hilbert action in this case.

While it should be possible to reconstruct the scalar curvature, and thus the Einstein Hilbert action, from many different measures on the causal set, one of the easiest ways to do so is to start from the d'Alembertian operator.
To define a discretized derivative on the causal set we need to rely purely on relativistically covariant information.
In this case we write a retarded d'Alembertian operator, at an element $x$ by using the past of the element.
In $2$ dimensions the expression is
\begin{align}
B^{(2)}\phi(x)&:=\frac{1}{l^2}\Big[- 2\phi(x) +4\Big({\sum_{y \in L_0(x)}\!\!\!\phi(y)}-
2\!\!\! {\sum_{y\in L_1(x)}\!\!\!\phi(y)}+{\sum_{y\in L_2(x)}\!\!\!\phi(y)}\Big)\Big]
\end{align}
where $L_i(y)$ are all elements $y$ of the causal set for which $|I_A(x,y)|=i$, with $I_A(x,y)$ the Alexandrov interval, consisting of all elements causally between $x,y$~\cite{Sorkin_2007,Benincasa_Dowker_2010}.
The pre-factor $\frac{1}{l^2}$ is for dimensional reasons, with $l^d$ the $d$-dimensional discreteness volume.
This is a sum over `layers' of the causal set, where using the Alexandrov intervals to define the layers makes these covariant.
The $2$d expression above can be generalized to arbitrary dimension~\cite{Dowker_Glaser_2013,Glaser_2014},
\begin{align}\label{eq:localdalembert}
B^{(d)}\phi(x)=\frac{1}{l^2}\left( \alpha_d \phi(x) +\beta_d \sum\limits_{i=0}^{n_d}  C^{(d)}_i \sum\limits_{y \in L_{i}}\phi(y) \right)
\end{align}
where  $\alpha_d$, $\beta_d$, $C_i^{(d)}$ and $n_d$ are dimension dependent constants.

To determine these constants we demand that this expression agrees with the d'Alembertian operator, when applied to a casual set sprinkled into flat Minkowski space.

For such a causal set the number of points in the $i$-th layer is:
\begin{align}
L_i(y)= \int\limits_{J^-(0)} \mathrm{d}V_y \frac{(l^{-d} V_{d}(y))^i}{i!} \,e^{-l^{-d} V_{d}(y)}
\end{align}
which we can insert above, to write down a continuum operator
\begin{align}
l^{2}(\square_l^{(D)} \phi)(x)=\alpha_d \phi(x)+ \sum_{i=0}^{L_{max}}\frac{\beta_d C_i^{(d)} }{i!} \int\limits_{J^{-}(x)} e^{-l^{-d} V(x,y)} \left( l^{-d} V(x,y)\right)^i \phi(y) \md V_y \;.
\end{align}
The constants $C_i^{(d)}$ can be generated through differential operators $\mc{O}_d$ in the discreteness scale $l$, which for even dimension are defined as $\mc{O}_{2n}= \frac{1}{2^{n+1} (n+1)!} (2- l\frac{\partial}{\partial l})(4- l\frac{\partial}{\partial l}) \dots (2n+2-l\frac{\partial}{\partial l})$ and $\mc{O}_{2n+1} = \mc{O}_{2n}$.
Using this continuum approximation we can calculate
\begin{align}
	\frac{1}{\beta_{d}}&= \lim_{l \rightarrow 0}  \frac{1}{2 l^{d+2}} \mathcal{O}_{d} \int \md V_y  \left( \frac{v_y-u_y}{\sqrt{2}} \right)^{2} e^{-l^{-d} V_{0\,d}(y) }  \\[10pt]
\frac{\alpha_d }{\beta_d}&= - \lim_{l \rightarrow 0}
\frac{1}{l^{d}}   {\cal{O}}_d \int \mathrm{d} V_y e^{-l^{-d} V_{0\,d}(u,v)}  \;.
\end{align}
This can be solved to find
\begin{align}
\alpha_d&= \begin{cases}
\frac{- 2 c_d^{\frac{2}{d}}}{\G{\frac{d+2}{d}}} & \text{for even }d \\
\frac{- c_d^{\frac{2}{d}}}{\G{\frac{d+2}{d}}} & \text{for odd }d\\ \end{cases} &
\beta_d&= \begin{cases}
\frac{2 \; \G{\frac{d}{2}+2} \G{\frac{d}{2}+1}}{\G{\frac{2}{d}} \G{d}} \; c_d^{\frac{2}{d}} & \text{for even }d\\
\frac{d+1}{2^{d-1}\G{\frac{2+d}{d}}} \; c_d^\frac{2}{d} & \text{for odd }d\\
 \end{cases}
\end{align}
where $c_d= S_{d-2} \frac{2^{\frac{2-d}{2}}}{d(d-1)}$ with $S_{d-2}$ the volume of the unit $d-2$ sphere.

\begin{figure}
  \centering
  \includegraphics[width=0.75\textwidth]{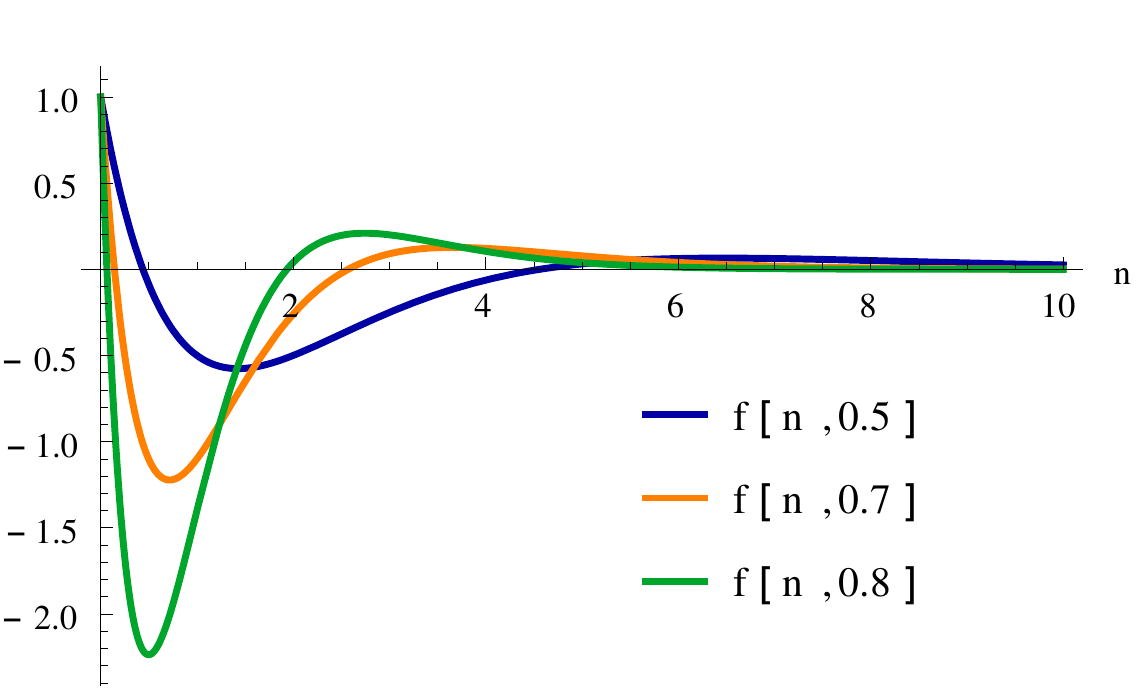}
  \caption{Plot of the smoothing function $f(n,\epsilon)$ for $2$d and different values of $\epsilon$ \label{fig:fne}}
\end{figure}
As Sorkin already realized in \cite{Sorkin_2007} this expression has the correct expectation value.
However, in the high density limit the fluctuations around this mean increase.
He then proposed to resolve this issue by introducing an intermediate non-locality scale, and defining, for each $d$, a one parameter family of operators on scalar fields on a causal set
\begin{equation}
B^{(d)}_\epsilon \phi(x):=\frac{\epsilon^{\frac{2}{d}}} {l^2}\left( \alpha_d \phi(x) +
\beta_d \epsilon \sum\limits_{y\prec x}f_d(I_A(x,y),\epsilon)\phi(y)\right) \, , \label{eq:nonlocdAlembertian}
\end{equation}
where the sum is over all elements $y$ in the causal set to the past of $x$ and
\begin{equation}\label{eq:lazybones}
f_d(n,\epsilon ):=(1-\epsilon)^n
\sum_{i = 1}^{n_d} C_i^{(d)} {n \choose i-1} \left(\frac{ \epsilon}{1-\epsilon}\right)^{i-1}\,,
\end{equation}
where $\epsilon=V_{Pl}/V$ where $V_{Pl}=l^d$ is the Planck volume and $V$ is the volume of the intermediate non-locality scale.
In the limit of $V\to V_{Pl}$ this agrees with equation \eqref{eq:localdalembert} above, which corresponds to the action with minimal non-locality.
For larger $V$/ smaller  $\epsilon$ the smearing function $f_d(n,\epsilon)$ becomes less local, and includes more layers, as illustrated in Figure~\ref{fig:fne}.
This suppresses the fluctuations of the operator.

To derive an expression for the scalar curvature from this, we generalize from calculating the d'Alembertian on Minkowski space, to a curved spacetime.
We can use a Riemann normal coordinate expansion, to derive that
\begin{align}\label{eq:boxricci}
\lim_{l \rightarrow 0}
{\bar{B}}^{(d)}\phi(x) &= \Box^{(d)} \phi(x) + \frac{1}{2} R(x)\phi(x)\,.
\end{align}
When applying this operator to a constant scalar field $\phi(x)=2$, the only contributing term is the scalar curvature at a given point.
To derive the Einstein-Hilbert action, we integrate this over spacetime, which in the case of the causal set is a sum over all causal set elements.
Assuming the discreteness is at the Planck scale $l=l_p$, this leads to the following expression \cite{Dowker_Glaser_2013},
\begin{align}
\frac{1}{\hbar} S_{dD} = \alpha_d N + \beta_d \sum_{i = 1}^{n_d} C_i ^{(d)} N_i \;,
\end{align}
for a $N$ element causal set, where the numbers $N_i$ are the abundances of $i$ intervals, which can be calculated analytically~\cite{Glaser_Surya_2013}.
In addition to their use in the action, these abundances are also useful as measures of manifold likeness of a causal set and as measures of locality for subregions of a causal set.

In the case of $d=2$, which is used in the simulations we will talk about, this leads to
\begin{align}\label{eq:2daction}
\phantom{ \sum_{i = 1}^{n_d}}\hspace{-9pt} \frac{1}{\hbar} S_{2D} = N- 2  N_0 +  4 N_1 - 2 N_2
\end{align}
for $\epsilon=1$ and
\begin{align}
  \phantom{ \sum_{i = 1}^{n_d}}\hspace{-9pt} \frac{1}{\hbar} S_{2D} (\epsilon) = 4 \epsilon \left(N - 2 \epsilon \sum_{n=0}^{N-2} N_n f_2(n,\epsilon)\right) \\
  f_2(n,\epsilon) = (1-\epsilon)^n \left(1- \frac{2 \epsilon n}{(1-\epsilon)}+ \frac{\epsilon^2 n (n-1)}{2 (1-\epsilon)^2}\right)
\end{align}
for arbitrary $\epsilon$.
In the computer simulations the non-locality parameter $\epsilon$ is a free input and thus defines a class of actions.

\subsection{Markov Chain Monte Carlo Simulations}\label{subsec:MCMC}
In a quantum theory, states do not have a statistical weight, instead they have a quantum amplitude.
The usual solution when applying simulations like this to quantum gravity systems is to introduce a Wick rotation, going from a quantum Lorentzian system to a statistical Euclidean system.
Since causal set theory is intrinsically Lorentzian, this is not possible, instead introduce an inverse temperature parameter $
\beta$ which we analytically continue $\beta \to i \bar{\beta}$
\begin{align}
  e^{i \beta S_{CST}} \to e^{ - \bar{\beta} S_{CST}}\;.
\end{align}
The parameter $\beta$ now becomes a new free parameter of our theory.
Some of the philosophy about why a parameter Wick rotation is useful in this context, has been developed in \cite{Sorkin_2012}

The studies described in this chapter are all done using Markov Chain Monte Carlo (MCMC) simulations, which are a tool for importance sampling from a large class of objects.
They are often  used in statistical physics applications, since they allow us to probe large sample spaces efficiently by favoring the regions that contribute most to the state sum.
They do so by sampling the state space through a Markov Chain, generated using an ergodic set of moves through state space.

An example of such a move, for our system of the $2$d orders, is the coordinate flip move, first constructed in~\cite{Surya_2012}.
It uses the $u_i,v_i$ labels of the orders, to propose a new order by flipping one of the coordinates.
\begin{enumerate}
  \item Randomly pick two of the $N$ elements $(u_1,v_1)$ and $(u_2,v_2)$ of the causal set.
  \item Randomly pick one of the two labels $a \in (u,v)$, corresponding to one of the two light-cone directions.
  \item The proposed new causal set is the $2$d order with the labels $a_1,a_2$ interchanged.
\end{enumerate}
This move is clearly ergodic in the space of $2$d orders, and easy to implement on the computer.
Code using this move to simulate the $2$d orders, together with an added matter coupling, as described below can be found in~\cite{code}.

The MCMC algorithm most often used in quantum gravity theories is Metropolis Hastings, as for example explained in~\cite{Newman_Barkema_1999}.
The algorithm starts at some configuration of the system $s_1$\footnote{The exact initial configuration is in theory irrelevant, since it can be proven that for long enough running times the algorithm will converge towards the correct sampling probability for all states.
In practice the choice of initial configuration can massively slow down the convergence of the algorithm, and ensuring that the algorithm has converged towards the correct sampling is a non-trivial task.
We will not go into more detail on this here, but refer again to the excellent textbook by Newman and Barkema~\cite{Newman_Barkema_1999}.}, from this initial state, the algorithm generates a new state, using the prescribed moves.
It is important that the set of moves is ergodic, to ensure that the entire space of states can be sampled.
After a new state is generated the algorithm will append either the new state, or an additional copy of the current state to the chain.
The transition probability between the current state $s_1$ and a new proposed state $s_2$ in the chain is given by their statistical weight.

If the new state has a higher statistical weight (lower energy) than the current state, it will always be accepted.
Even if the new state has a lower statistical weight (higher energy) it has a chance of being accepted, with probability $exp(S_{\text{old state}}-S_{\text{new state}})$.
This is important, to avoid the chain getting stuck in a state that is a local minimum, in relation to the states that are adjacent to it under the set of moves, but which is not a global minimum.

\section{Simulations of $2$d orders}\label{sec:sims}
This section will study three different explorations of the $2$d orders using MCMC simulations.
The first subsection~\ref{subsec:2d} is about the phase diagram of the $2$d orders, their remarkably simple scaling behavior, and their first order phase transition.
The second~\ref{subsec:HH} is about an attempt to use the $2$d orders to study a toy model of a wave function of the universe, transitioning from nothing to fixed size in the causal set, and the last~\ref{subsec:Ising} explores how the phase structure of the $2$d orders changes when they are coupled to an Ising model.

\subsection{A first order phase transition}\label{subsec:2d}
\begin{figure}
  \centering
\includegraphics[width=\textwidth]{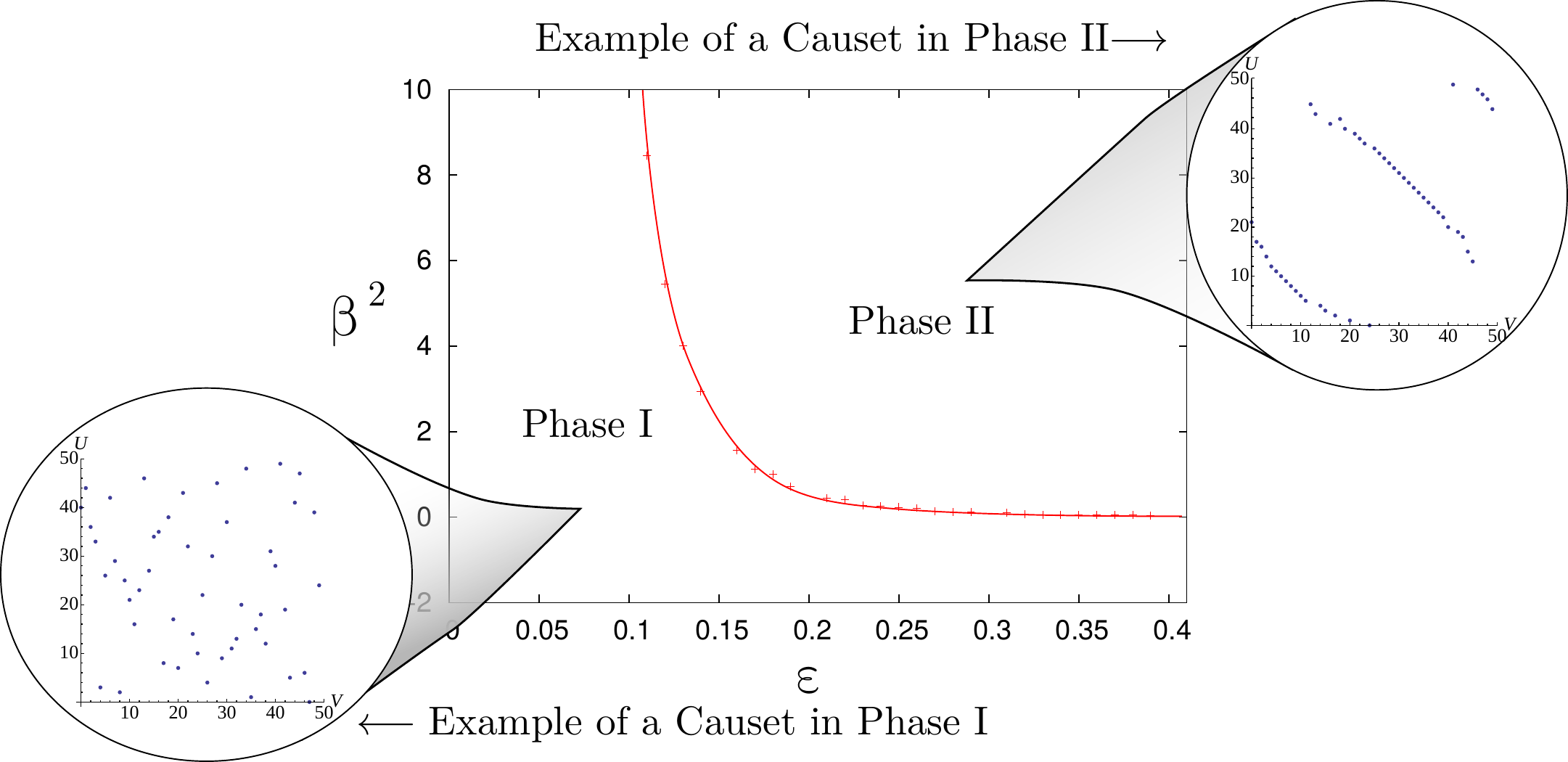}
  \caption{
  Phase diagram for the $2$d orders in the $\epsilon, \beta^2$ plane for $N=50$.
  The insets show a typical causal set in either of the two phases, to illustrate the dominant contribution.
  Plot taken from~\cite{Surya_2012} with slight modification to include images of the two phases.
  }
   \label{fig:2dptoriginal}
\end{figure}

The first work on simulations of the $2$d orders was done by Sumati Surya~\cite{Surya_2012}.
Careful study revealed that, using the action~\eqref{eq:2daction}, the $2$d orders show two distinct phases.
Simulations for different values of $\epsilon$ and $\beta$ make it possible to trace the phase transition line shown for $N=50$ in Figure~\ref{fig:2dptoriginal}.

The two phases are the random $2$d orders and the crystalline orders, illustrated through insets in the figure.
The low $\beta$ phase is dominated by the random $2$d orders, which are the entropically favored states, while high $\beta$ energetically favors the creation of layers.
The energetically optimal state would be a set with two layers, thus maximizing the number of links, however since this configuration exists exactly once it is entropically strongly disfavored.

To delineate the different phases a number of observables were used, the number of links, the value of the action, the height of the $2$d order, the ordering fraction, time asymmetry and the interval abundance~\cite{Glaser_Surya_2013}.
The height is defined as the length of the longest chain, and the ordering fraction is the number of links, divided by the maximal number of links possible in a set of this size (which is $N(N-1)/2$, achieved exactly if the causal set consists of two equally sized layers where each element is connected to all elements in the other layer).
The time asymmetry is measured by proxy, as the difference between the number of maximal and minimal elements in the causal set.
The action of the system is time reversal invariant, it is thus interesting to see whether the dynamics of the system will break this invariance.

The interval abundance is a particularly interesting observable, that was first introduced in~\cite{Surya_2012} and then studied in depth, analytically as well as numerically, in~\cite{Glaser_Surya_2013}.
The interval abundance $I_A(n)$ counts the number of Alexandrov Intervals of  size $n$. 
This defines a curve that can be calculated exactly for flat space, and that contains a lot of information about the causal set.
In particular, it can recognize the dimension of a region, and recognize whether it is a locally flat region in a larger curved causal set.
It can also be calculated for a slightly curved space-times~\cite{Machet_2020}.

Studying these quantities shows that the phase of random $2$d orders has an ordering fraction of $0.5$, an action of $\sim 4$ a height of $~10$ and no time asymmetry.
All of these observations are in good agreement with those for a causal set generated through sprinkling into $2$d Minkowski space, and thus confirm that this is a continuum phase.
This can also be tested using the interval abundance as a measure of manifoldlikeness~\cite{Glaser_Surya_2013}.

The crystalline $2$d orders on the other hand show clear non-manifoldlike behavior.
They are energetically favored states with a large number of links, and are marked by a relative high ordering fraction $\sim 0.6$, a small action $\sim -45$ a low height of just $3$ and large time asymmetry, although the direction of the time asymmetry fluctuates.

The existence of these two phases opens the question of what order the transition between them is.
To study the order of a phase transition one needs to perform a scaling analysis, comparing the behavior at the phase transition for different system sizes.
This, together with a thorough analysis to understand how the system scales for varying in $\epsilon$, was undertaken in~\cite{Glaser_O’Connor_Surya_2018}.

To explore the scaling, simulations were done for $N$ between $30$ and $90$ and for $\epsilon$ from $0.1$ to $0.5$, covering an irregular grid of these values.
The range for $\epsilon$  was chosen such since low and high $\epsilon$ are problematic.
For very low values of $\epsilon $ the system shows an additional phase, which was not studied in detail, since for very low $\epsilon$ at fixed size $N$ the scale of the smearing of the action becomes much larger than the size of the set.
For very high $\epsilon$ the fluctuations in the action become strong, which requires much longer simulations to study.

\begin{figure}
\subfloat[Histogram demonstrating coexistence]{\includegraphics[width=0.5\textwidth]{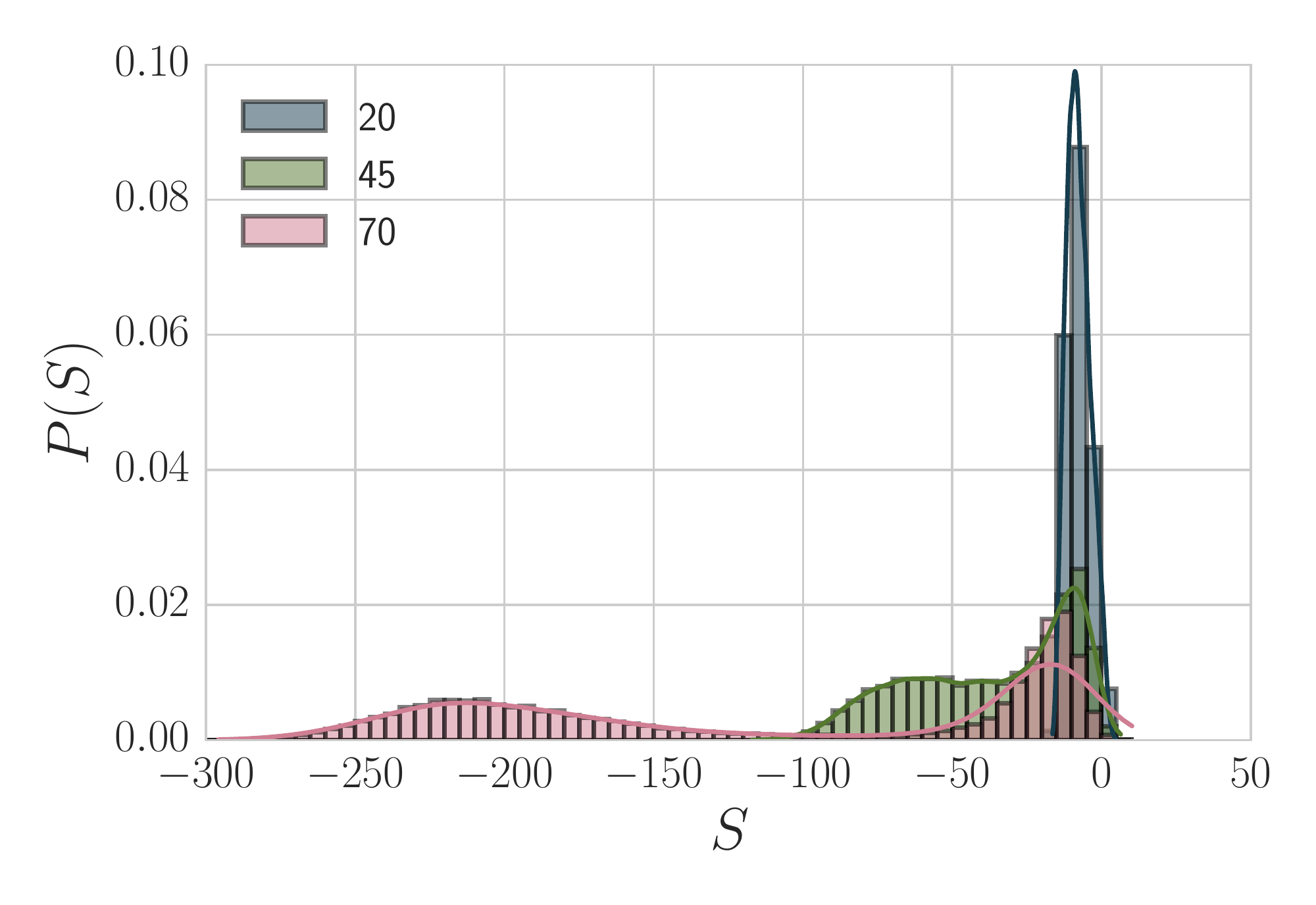}}
\subfloat[Binder coefficient]{\includegraphics[width=0.5\textwidth]{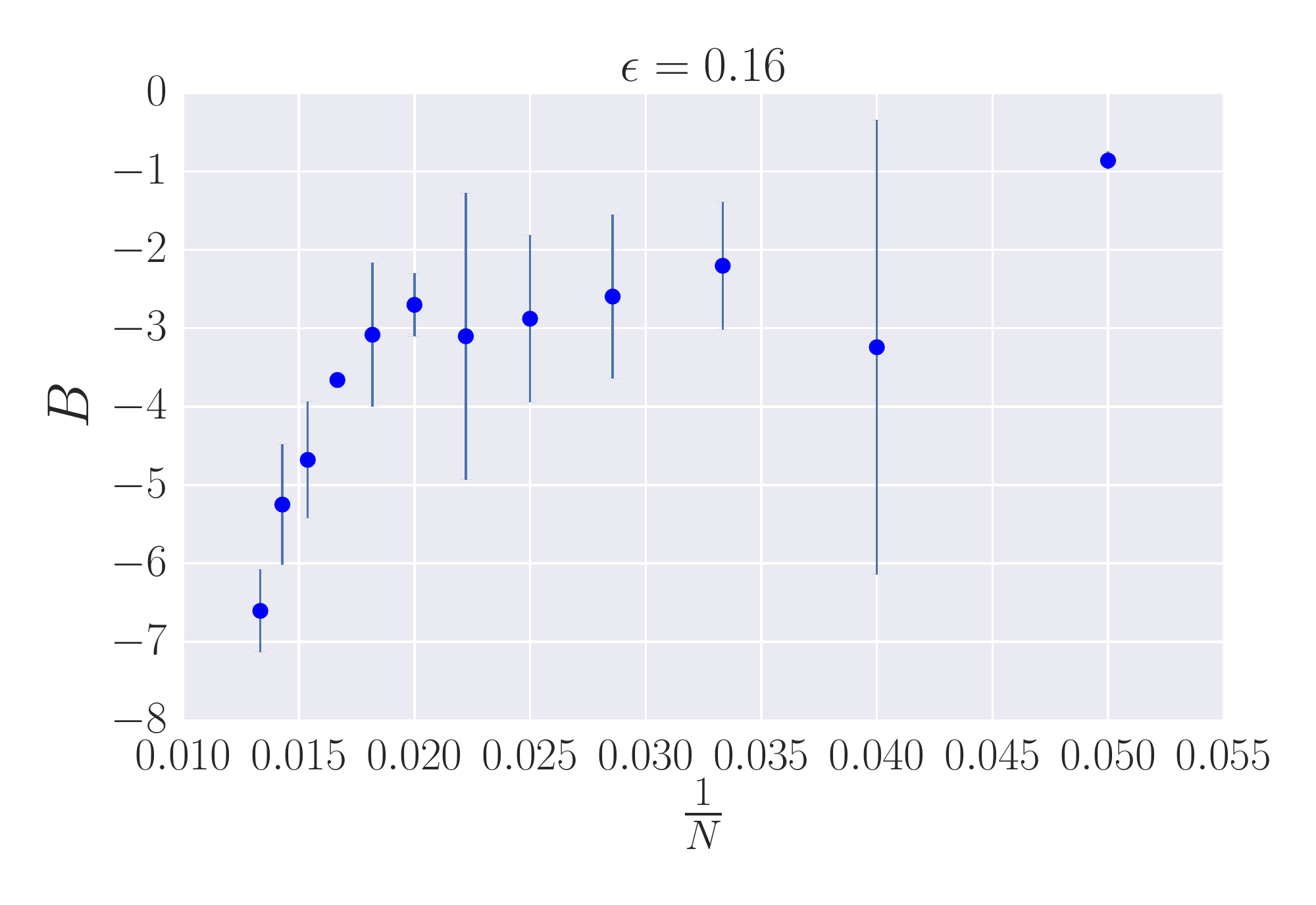}}
  \caption{The left-hand figure shows a histogram, depicting the occurrence of different action values in the simulations, showing that the system fluctuates between two different states, which become more different as the system size increases.
  This indication of a first order phase transition is confirmed by the right-hand plot, which shows the Binder coefficient plotted against $1/N$.}
  \label{fig:determinethePT}
\end{figure}

The first question explored was the type of the phase transition, which was determined to be of first order.
This was done through exploration of the histogram of the action, and using the Binder coefficient~\cite{Binder_1981}.
The histograms are shown in Figure~\ref{fig:determinethePT} (a).
The important observation here is that as $N$ increases the histogram at the phase transition splits further apart.
At a higher order phase transition the system at the phase transition point would show new, long range correlated, behavior, and thus form a single peak as $N$ increases.
The Binder coefficient is defined as
\begin{align}
  B=\frac{1}{3} \left( 1- \frac{\av{S^4}}{\av{S^2}^2}\right)
\end{align}
and is a useful tool to observe the order of a phase transition, exactly because of the behavior described above.
For a higher order phase transition, this coefficient should go towards $0$ as $N\to \infty$, the coefficient is shown in Figure~\ref{fig:determinethePT} (b).

It is thus clear that the $2$d orders do not undergo a higher order phase transition.
The existence of a higher order phase transition in a statistical system implies an infinite correlation length, and thus the existence of a continuum limit.
Since causal sets are a fundamentally discrete theory, the lack of a higher order transition is not condemning in itself, as a continuum limit is explicitly not expected.
It does however open interesting questions when observed from the vantage point of renormalization group techniques, where a higher order phase transition implies a universality of phenomena, which ensures that the fine structure at the Planck scale does not unduly influence physics at lower energies.
Some first steps towards renormalization group techniques for causal sets are made in~\cite{Eichhorn_2018}.
The lack of such a transition would then imply that there might be no universality, which could lead to problems in the predictivity of causal sets.
However, self averaging behavior, or other phenomena might resolve this difficulty, alternatively a more full description of causal sets might find a higher order transition after all.
\begin{figure}
  \centering
  \includegraphics[width=0.7\textwidth]{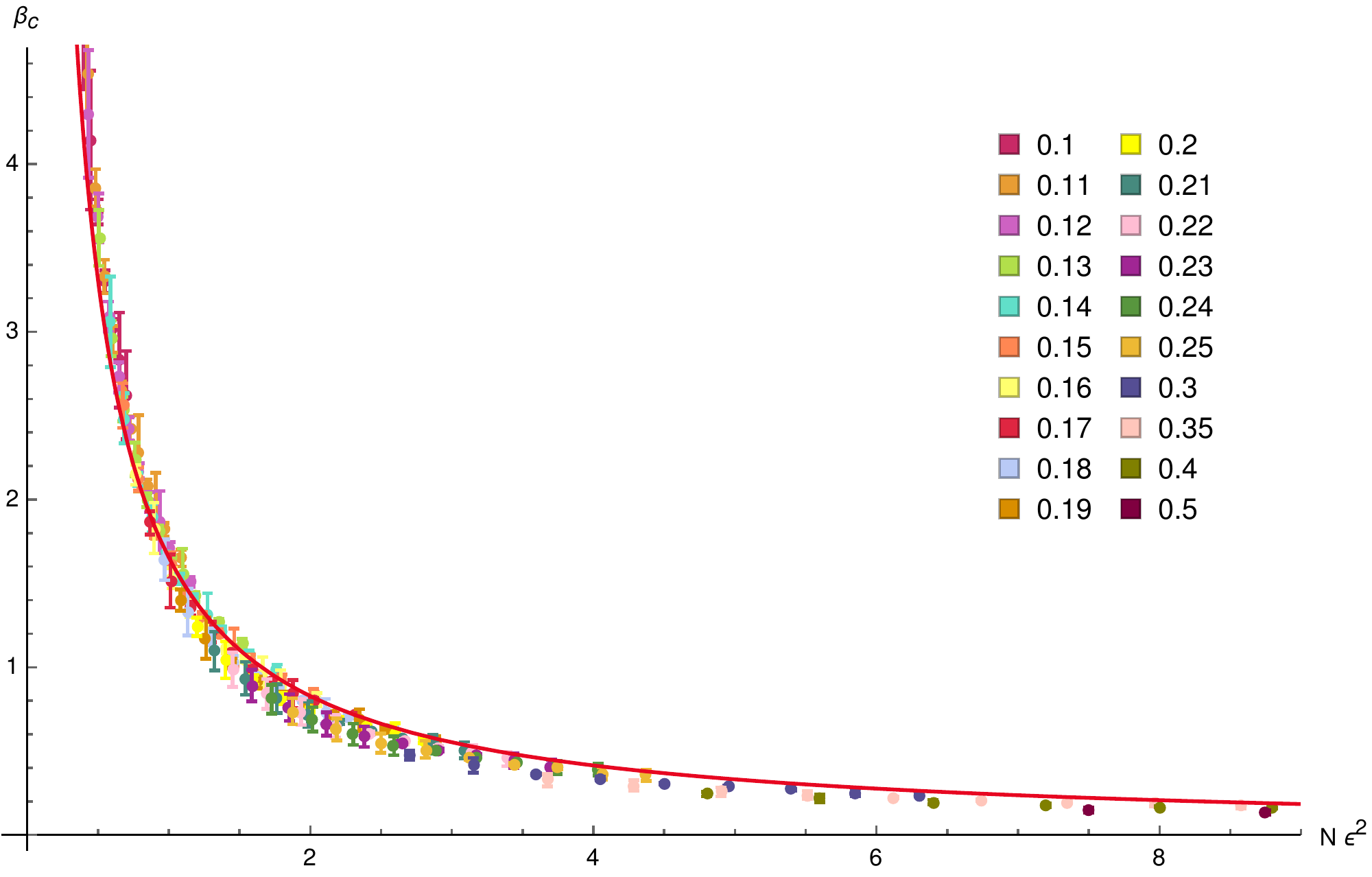}
  \caption{Plot of the phase transition point $\beta_c$ for all values of $\epsilon$ plotted against $N \epsilon^2$ to show that the scaling is an excellent description. The red line shows the best fit scaling. }
  \label{fig:PTscalingallfit}
\end{figure}
As shown in Figure~\ref{fig:PTscalingallfit} the location of the phase transition $\beta_c$ scales with $N \epsilon^2$.
In~\cite{Glaser_O’Connor_Surya_2018} this scaling was determined to be
\begin{align*}
	\beta_c(N,\epsilon) = \frac{1.66^{(\pm 0.03)}}{N\epsilon^2}
\end{align*}
plus some corrections in $N^{-2}$.

\begin{figure}
\subfloat[Scaling of $S$ $\beta < \beta_c $]{\includegraphics[width=0.5\textwidth]{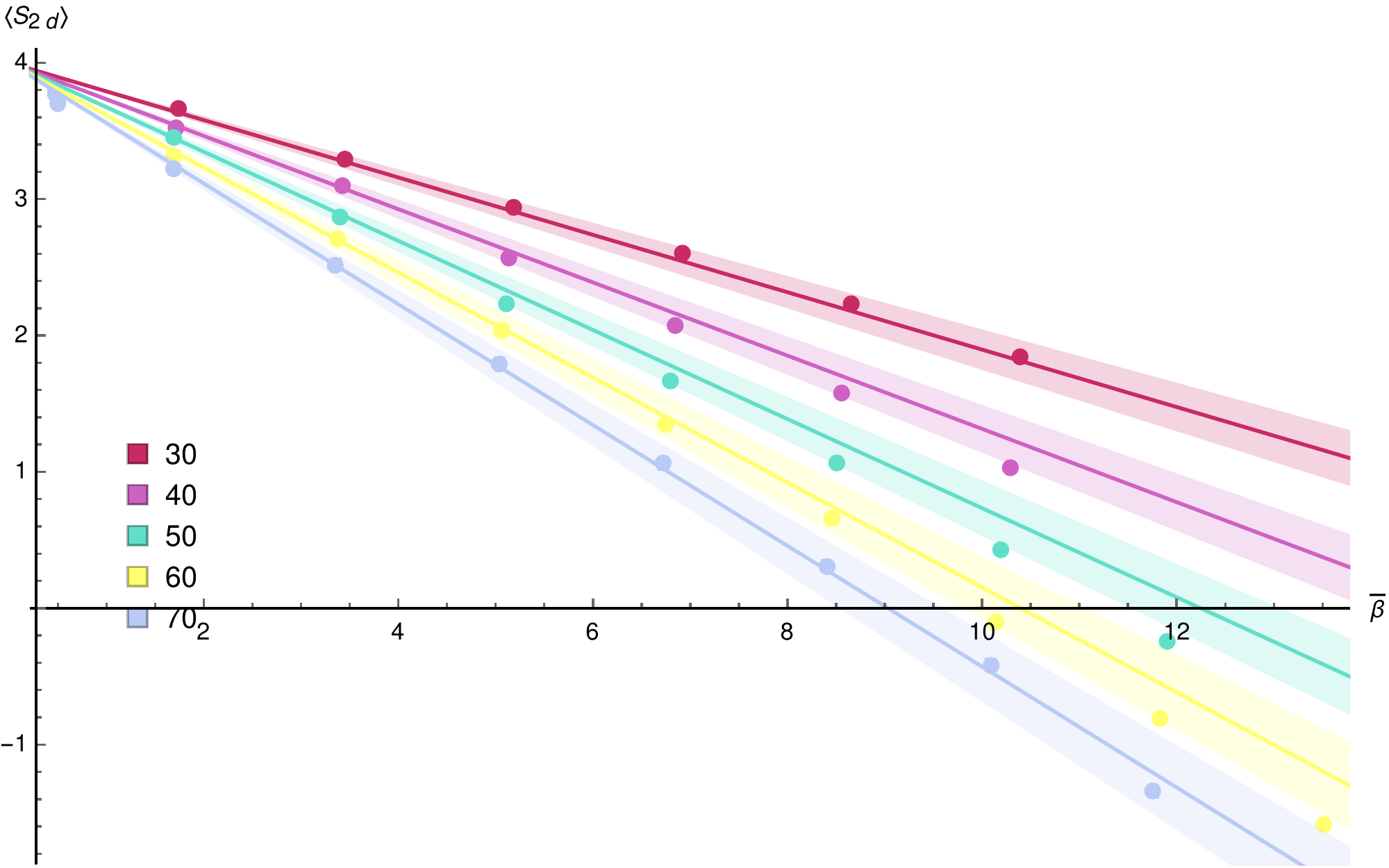}}
\subfloat[Scaling of $S$ $\beta> \beta_c$]{\includegraphics[width=0.5 \textwidth]{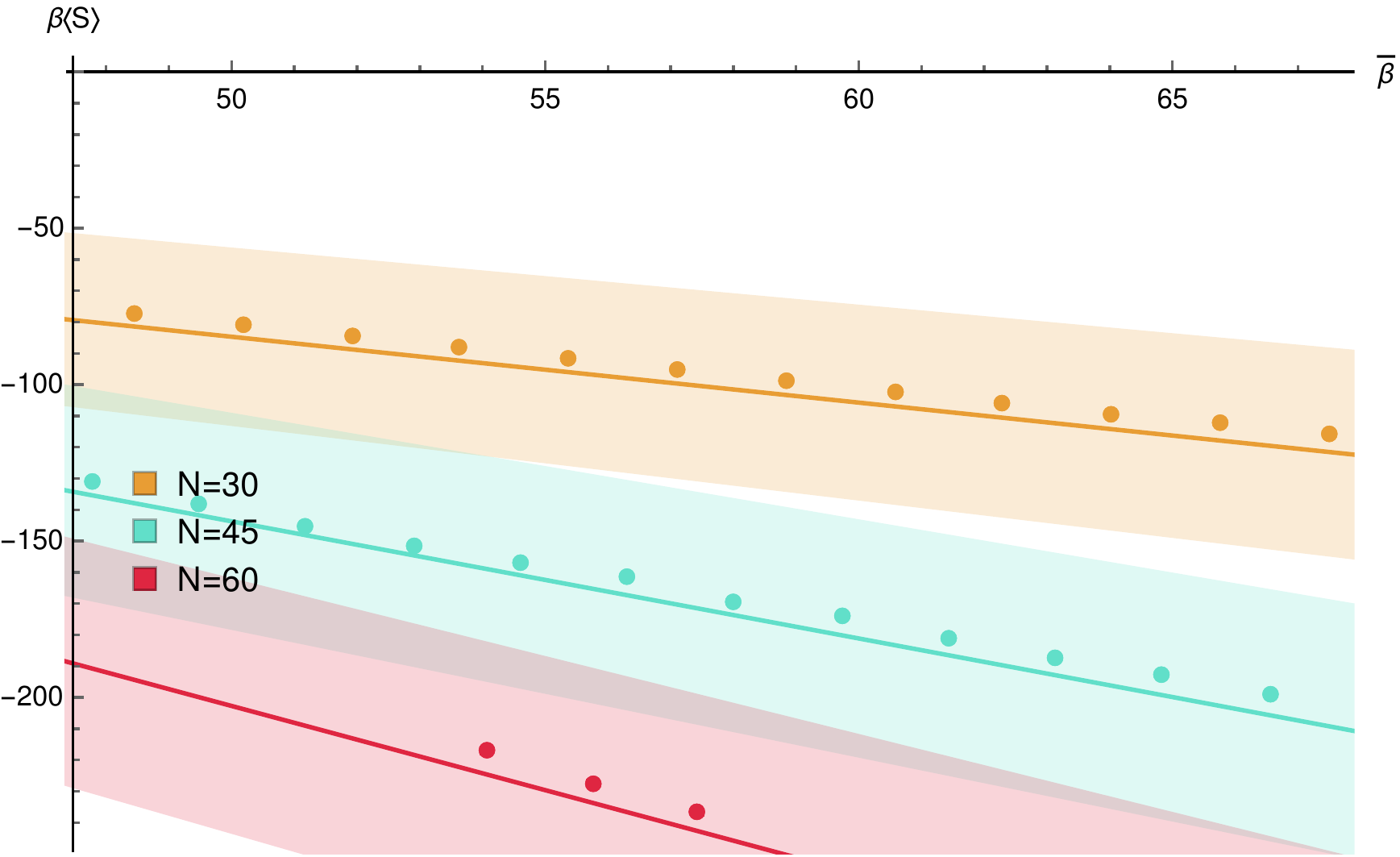}}

  \caption{
The left-hand side shows the average action plotted against $\bb$ for the region $\beta < \beta_c$, and the right-hand side shows this for $\beta>\beta_c$.
The straight lines are best fits and show that the action scales linearly in $N$ for both regions. This scaling remains for all $\epsilon$ with the scaling parameters simple functions of $\epsilon$.}
    \label{fig:linerascalingAction}
\end{figure}
Not only does the location of the phase transition scale well, but the value of the action also shows very clear scaling behavior.
To examine this behavior we plot the action against the scale invariant temperature $\bb= \beta N$.
The action scales linearly with $N$ for both the regions $\bb<\bb_c$ and $\bb>\bb_c$, as illustrated in Figure~\ref{fig:linerascalingAction}.

For $\bb<\bb_c$ we determined the scaling as
\begin{align}
  \Sa^{-}-4 & = (b^{-}_0(\epsilon) + b^{-}_1(\epsilon)N )\bar{\beta} \\
      b^{-}_0(\epsilon) &=2.09^{(\pm 0.55)}(\epsilon - 0.07^{(\pm 0.01)})^2 -190.50^{(\pm 15.46)}(\epsilon-0.07^{(\pm0.01)})^4\\
    b^{-}_1(\epsilon)&= -0.20^{(\pm 0.06)}\epsilon^3 - 2.04^{(\pm0.18)} \epsilon^2
\end{align}
this remains consistent for all $N,\epsilon$ combinations.
The subtraction of $4$ from the action makes sense since the expected value for this action on a flat Alexandrov interval is $4$, which is due to boundary contributions as examined in~\cite{Benincasa_2011,Buck_Dowker_Jubb_Surya_2015}.

In the region $\bb>\bb_c$ the scaling is
\begin{align}
  \beta \Sa^{+} &= a^{+}(N,\epsilon) + b^{+}(N,\epsilon) \bar{\beta}\\
  a^{+}&= -25.21^{(\pm 8.16)} + 1.53 ^{(\pm 0.21)} N \\
  b^{+}&= 0.4^{(\pm 0.24)} + 17.63^{(\pm 4.76)} \epsilon^2 - 2.48^{(\pm0.09)} \epsilon^2 N
 \end{align}
this is consistent with analytic insights one can gain from examining the action on the bilayer orders.

These fits work well for the action in their respective regions, and we can use finite size scaling to derive scalings for the variance from this, which all confirm the results.
We can thus say that we do understand the scaling of the action at the phase transition of the $2$d orders very well.
\subsection{The wave function of a universe}\label{subsec:HH}
MCMC simulations of the $2$d orders can calculate the expectation value of different observables, and thus show the phase structure of the theory.
In a quantum theory a typical observation would be a wave function, showing the transition probability from one state to another.
In quantum gravity this is exemplified by the Hartle Hawking wave function of the universe~\cite{Hartle_Hawking_1983}.
The ground state wave function over closed $3$ geometries $(\Sigma,h)$ is the Euclidean functional integral over $4$ geometries without an initial boundary
\begin{align}
 \Psi_0(\Sigma,h)= A \sum_{M} \int \md g^{E} e^{- I_E(g)}
\end{align}
where $\partial M = \Sigma ,g|_{\Sigma}= h $ and $I_E(g)$ is the Euclidean Einstein action.
This describes the transition from the no-boundary proposal to a $3$ geometry $(\Sigma, h)$.

To establish an equivalent of the HH wave function in a causal set, we need to consider the boundaries;
How does a final spatial boundary look in a causal set, and what is the equivalent of no boundary?
The simplest proposal for the final boundary is to fix the size of the longest anti-chain without future elements, the final slice of the universe if you will.
While, as an anti-chain, this does not directly contain any information about the geometry other than size, the geometric information is encoded in the connectivity to the causal set elements in the past, as shown in~\cite{Major_Rideout_Surya_2009}.
The closes analogue to the no boundary proposal is a single element as an initial condition.
This is the choice of initial state that matches up best with the picture of the universe arising from a single point, as described in~\cite{Glaser_Surya_2016}.
Using this we can define the HH wave function in CST as
\begin{align}
  \Psi_0 ^{(N)}(\mc{N}_f,\beta)& = A \sum_{C \in \Omega_{N}} e^{\frac{\beta}{\hbar} S(C)}\; ,
\end{align}
where here $\Omega_N$ is the set of $N$ element causal sets with a single initial element and $\mathcal{N}_f$ final elements.
In a first step we then restrict ourselves to only examine this wave function for the $2$d orders.

This is calculated in two steps, first MCMC simulations of the space of all $2$d orders with $N$ elements supplemented with analytic methods to obtain an estimate for the partition function $\mc{Z}_0(\mc{N}_f)$ of the restricted sets to use for the normalization.
Then additional simulations are done to obtain $\av{S_{2d}}_{\beta}(\mc{N}_f)$, which makes it possible to use numerical integration to calculate the wave function as
\begin{align}\label{eq:psi0}
      \Psi_0 ^{(N)}(\mc{N}_f,\beta)  &= A \mc{Z}_\beta(\mc{N}_f)= A \mc{Z}_{0}(\mc{N}_f) exp^{- \int_0^\beta \av{S_{2d}}_{\beta'1}(\mc{N}_f) \md \beta'} \;.
\end{align}
The normalization uses the partition function for $\beta=0$ $\mc{Z}_0(\mc{N}_f)$ and can, up to an irrelevant normalization, be obtained by MCMC simulations.
A problem in this approach is that the larger $\mc{N}_f$ becomes, the fewer configurations with this state exist.

\begin{figure}
  \centering
  \includegraphics[width=0.8\textwidth]{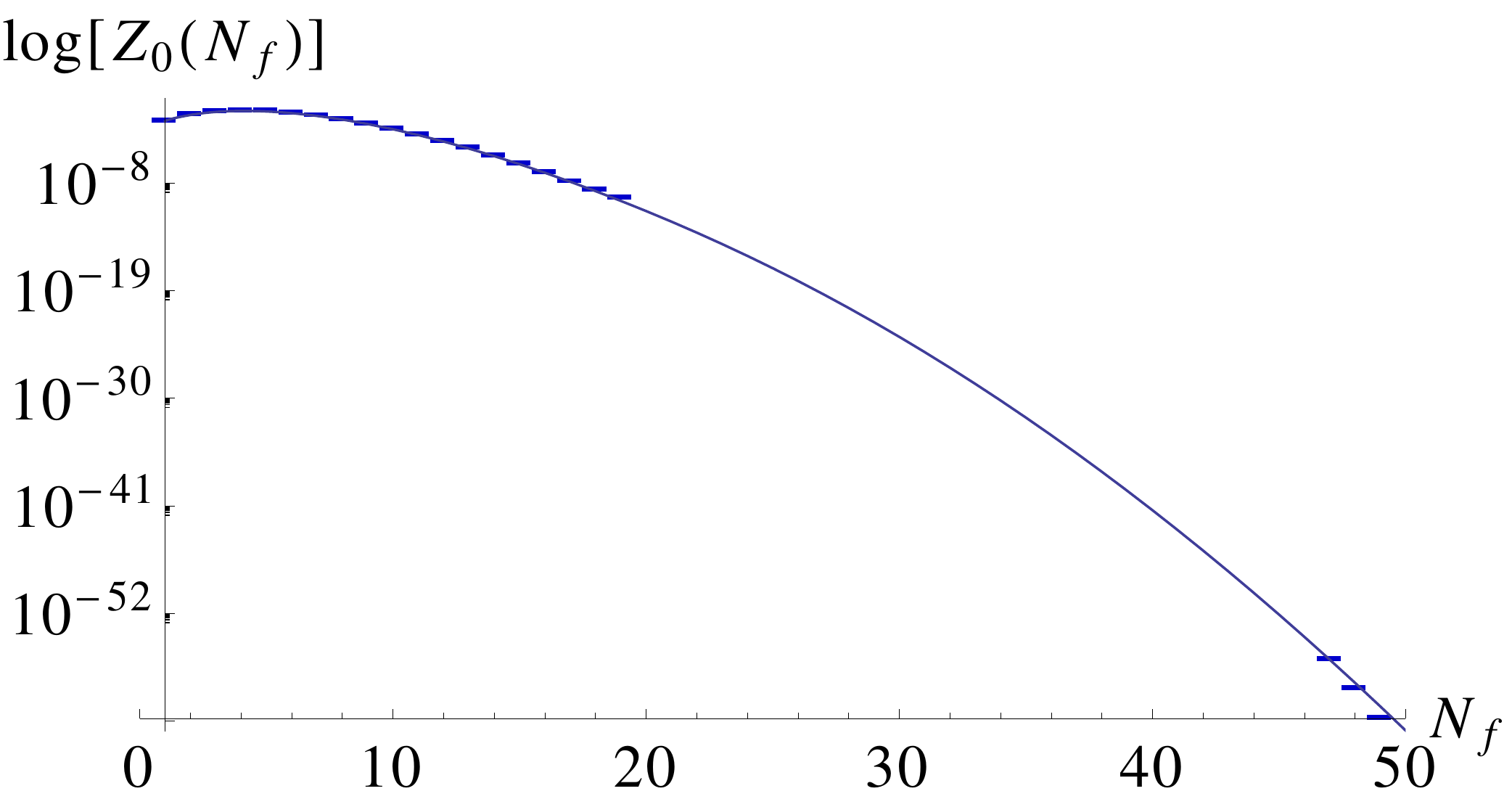}
  \caption{To determine the normalization factor $\mc{Z}_0(\mc{N}_f)$ for arbitrary values of $\mc{N}_f$ we generated a sample of configurations and counted the number of occurrences of different $\mc{N}_f$ values. To improve the fit we also added values for very large $\mc{N}_f$ which were calculated via combinatorial means.}
  \label{fig:Z0fig}
\end{figure}

In~\cite{Glaser_Surya_2016} the sets with $N=50$ are explored in depth, since this size has proven a good trade-off between manageable system size and reasonable confidence in our results, in past work.
We simulated $1.138\cdot 10^{10}$ $2$d random orders, which led to a sample of $2$d orders with final spatial boundaries of up to $\mc{N}_f=19$.
We were also capable to calculate the exact number of configurations arising for $\mc{N}_f=N-1,N-2,N-3$, thus giving us valuable additional information.
To obtain a value of $\mc{Z}_0$ for all values of $\mc{N}_f$ we then used a numerical fit, as shown in Figure~\ref{fig:Z0fig}.
The three analytically obtained points for $\mc{N}_f=49,48,47$ are important to anchor this fit, which is used to interpolate intermediate points.

\begin{figure}
  \centering
  \includegraphics[width=0.8\textwidth]{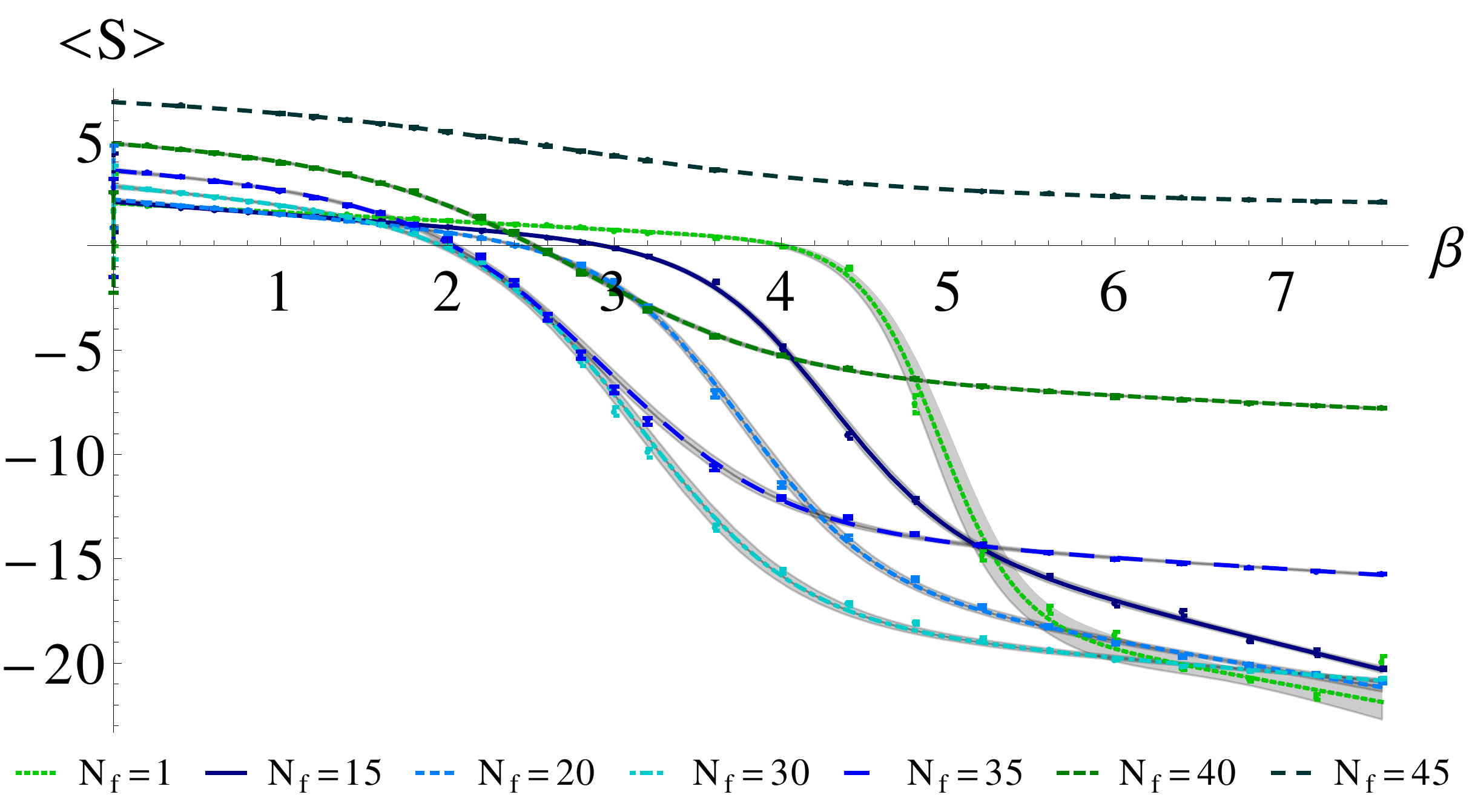}
  \caption{Average action for a range of $\mc{N}_f$ values for  $\epsilon=0.12$ with their best fit function and shading to show the uncertainty in the fit. }
\label{fig:avS}
\end{figure}

After thus having determined $\mc{Z}_0(\mc{N}_f)$ we need to determine  $\av{S_{2d}}_{\beta}(\mc{N}_f)$ for a range of $\mc{N}_f$ and $\beta$.
For this we ran separate simulations for each $\mc{N}_f,\beta$ combination.
We also explored different values of $\epsilon$, with the most extensive simulations for $\epsilon=0.12$ and some lower resolution simulations for $\epsilon=0.5, 1$ to check the consistency of our results.
In Figure~\ref{fig:avS} we show the data points for a selection of $\mc{N}_f$ plotted against $\beta$, with the best fit lines and their error regions.
The fit is done to obtain a smooth interpolation of the data present, which is necessary to be able to do the integral in~\eqref{eq:psi0}.
In theory splines or other piece wise interpolation methods lead to equivalent results.

\begin{figure}
  \subfloat[$\epsilon=0.12$]{\includegraphics[width=0.5\textwidth]{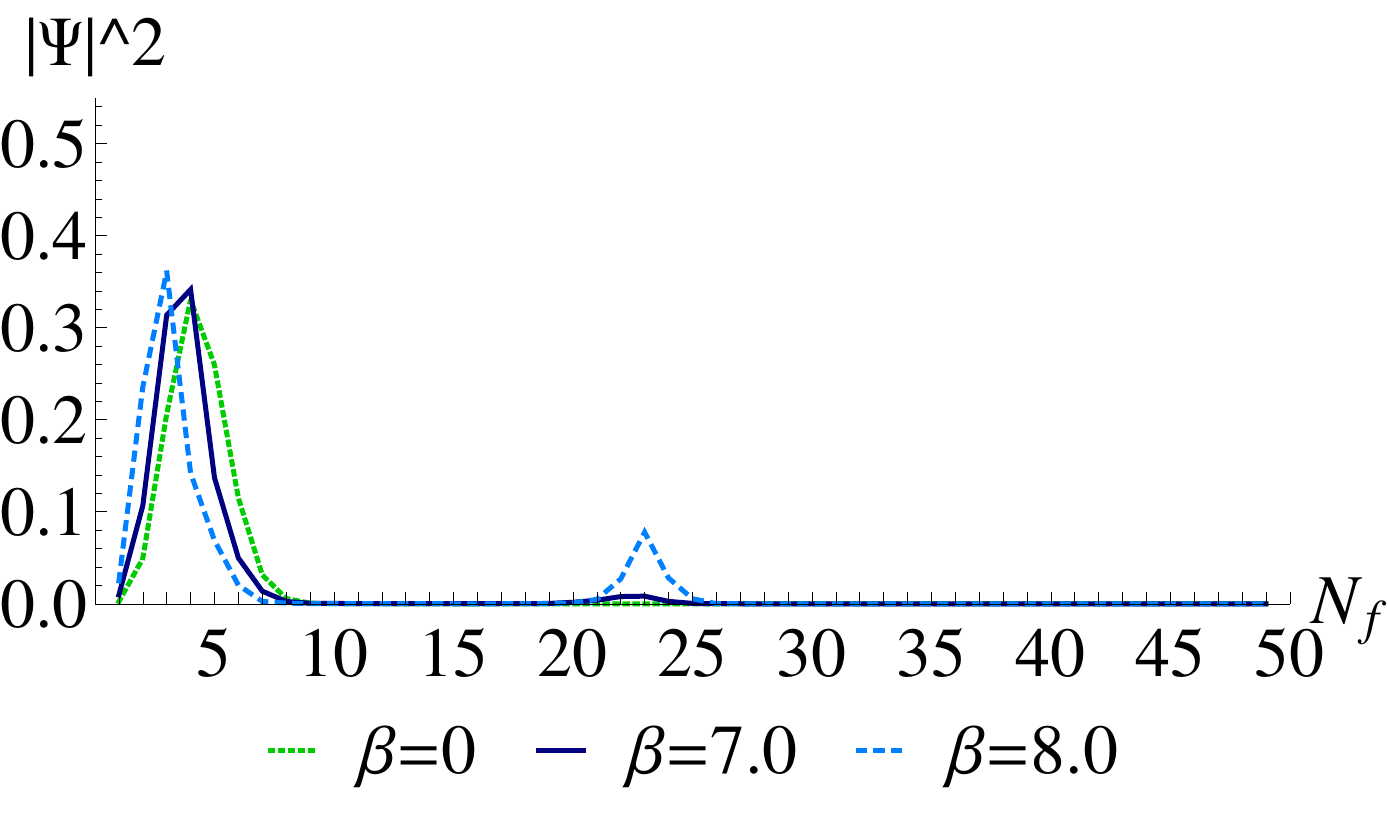}}
  \subfloat[$\epsilon=0.5$]{\includegraphics[width=0.5\textwidth]{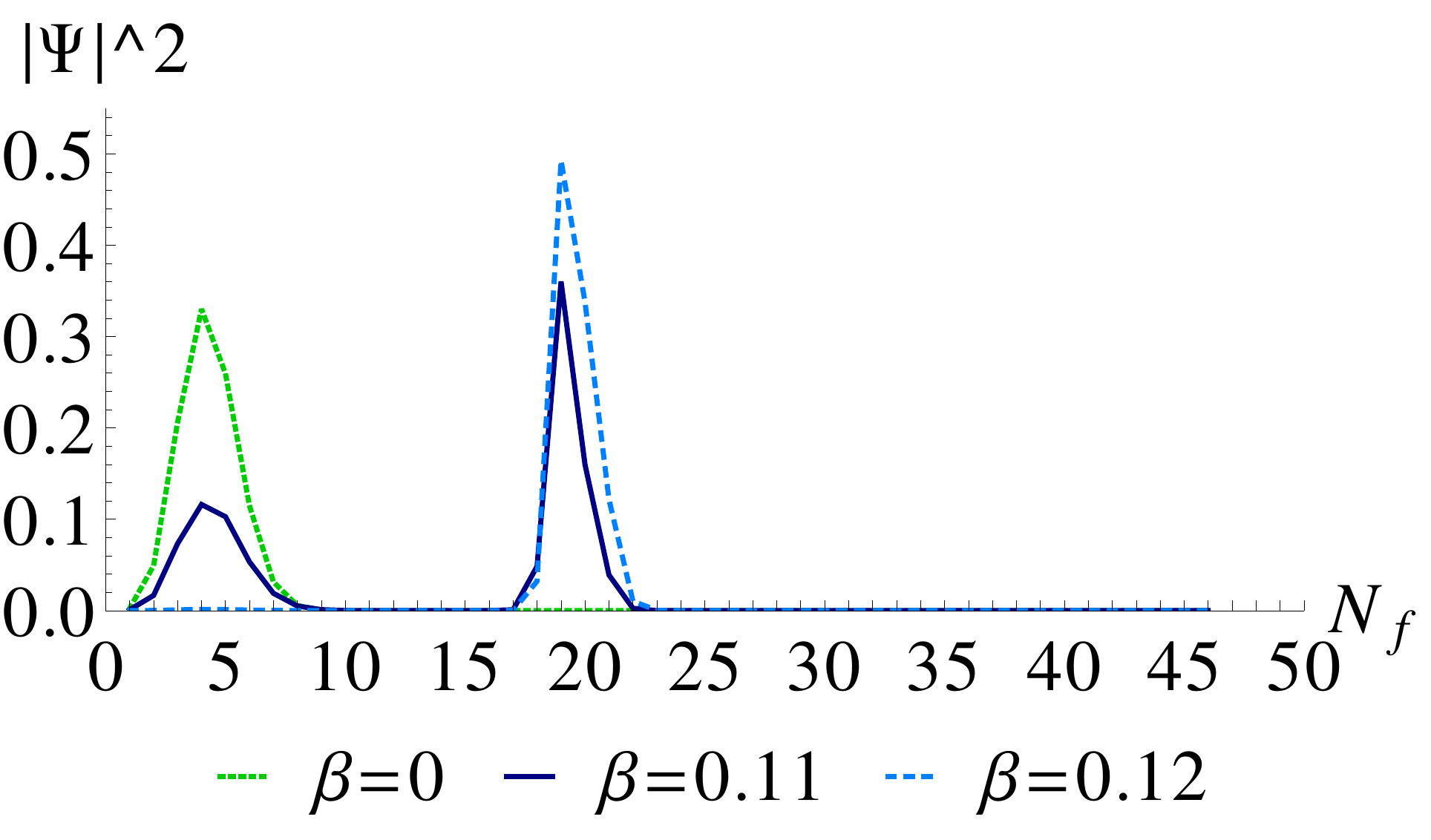}}
  \caption{Wave function for different values of $\beta$, showing how the wave function is concentrated around two peaks, one of which is dominant for small $\beta$ while the other becomes dominant at larger $\beta$}
  \label{fig:psi2}
\end{figure}

Using this function we numerically integrated, utilizing Mathematica, to obtain numerical values for $\Psi_0(\mc{N}_f,\beta)$.
These are shown in Figure~\ref{fig:psi2}, where the wave function of the universe is peaked around different final configurations for different $\beta$.

\begin{figure}
  \subfloat[$\mc{N}_f=4,\beta=0.2$]{\includegraphics[width=0.33\textwidth,angle=45]{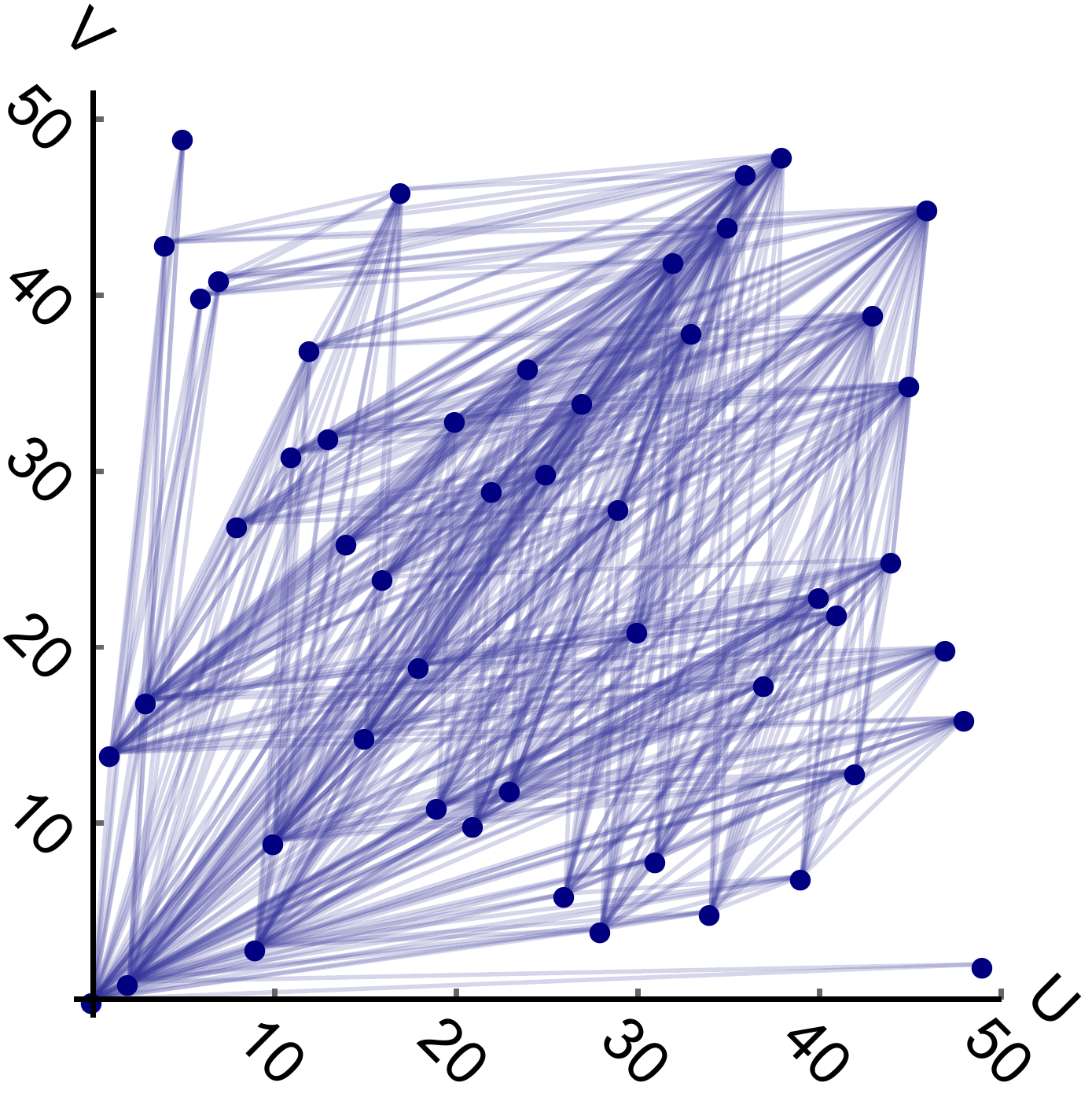}}
  \subfloat[$\mc{N}_f=23,\beta=7.6$]{\includegraphics[width=0.33\textwidth,angle=45]{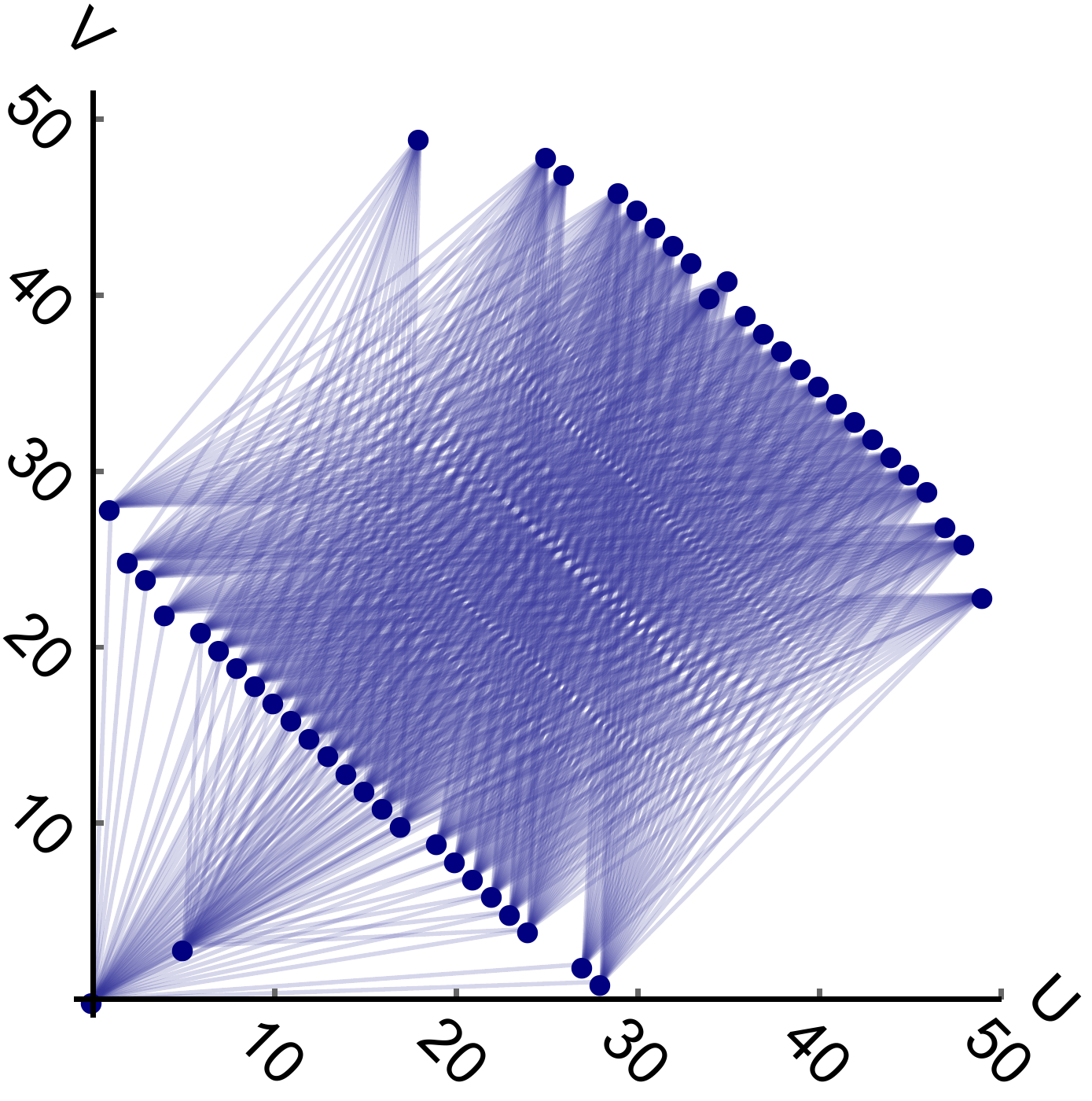}}
  \caption{Typical causal sets taken from the two peaks in the wave function for $\epsilon=0.12$. For low $\beta$ the typical configuration is random, while the typical configuration for large $\beta$ shows rapid expansion, and a crystalline behavior.}
\label{fig:excausets}
\end{figure}

The first peak is around $\mc{N}_f\sim4$, and is dominant for low $\beta$ and consists of the random, manifoldlike $2$d orders, which we already observed above.
The second peak is around $\mc{N}_f\sim 23$ and dominates once $\beta$ becomes high enough.
We show examples of typical causal sets in either of these peaks in Figure~\ref{fig:excausets}.

\begin{figure}
  \centering
  \includegraphics[width=0.6\textwidth]{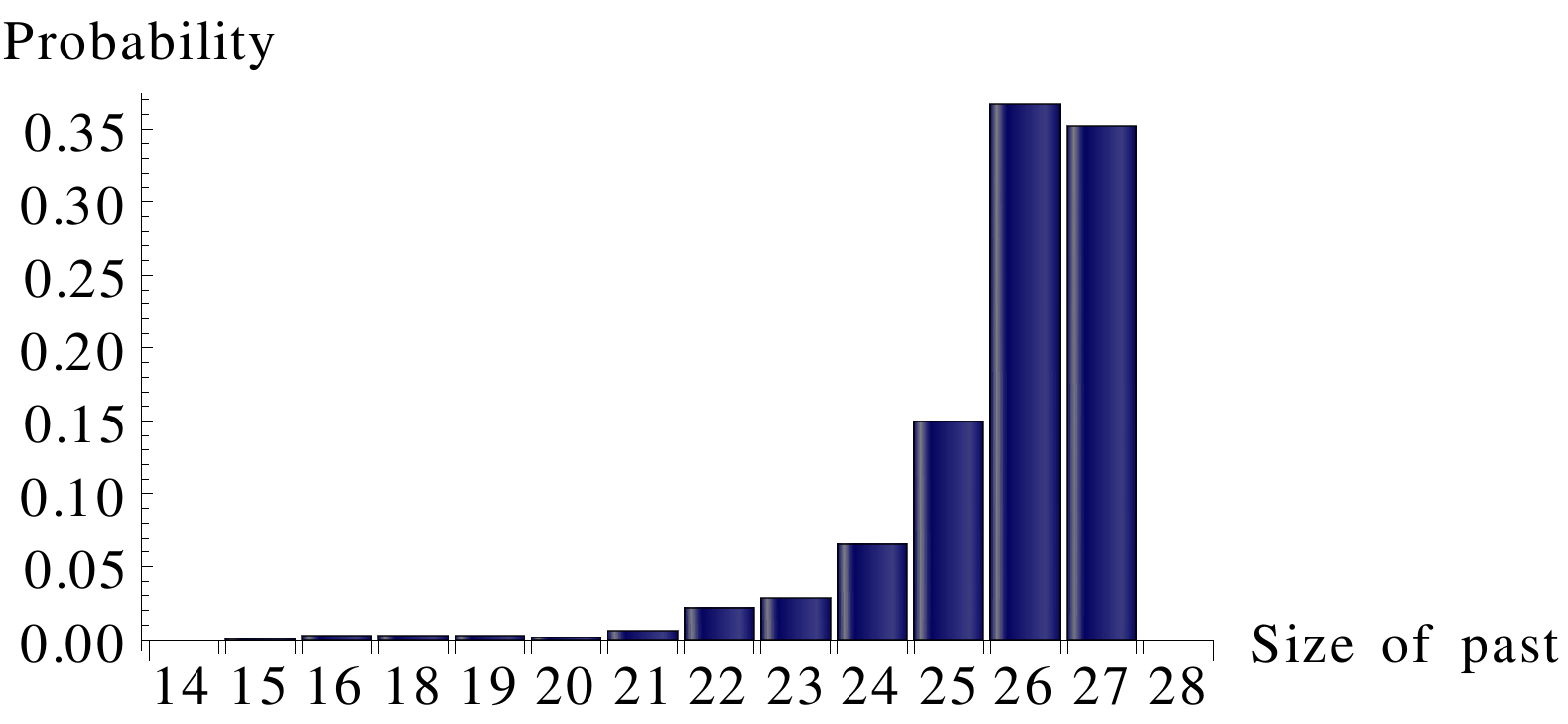}
  \caption{Probability histogram for the size of the past of a final element in the large $\beta$ dominant state. This shows that most final elements contain all bulk elements in their past.}
\label{fig:connectivity}
\end{figure}

The causal sets in the $2$nd peak are similar to those in the crystalline phase, but restricted by our initial condition of $\mc{N}_i=1$.
These configurations show several properties that are physically interesting.
They are rapidly expanding, as determined by the ratio of the largest antichain to the longest chain in the set.
It would be interesting to see how this would continue for larger $N$.
Another suggestive property of these causal sets is the distribution of pasts, as shown in Figure~\ref{fig:connectivity}.
The histogram shows the average distribution of past volumes for the final elements for $\mc{N}_f=23, \beta=7.6$.
On average each of the final elements is connected to almost all other elements in the causal set.
This high connectivity leads to a very homogeneous initial condition on this slice of the universe, similar to what is also achieved through inflation.
This is an interesting hint that discreteness in the early universe could give rise to the homogeneity of the universe.

\subsection{Adding matter -- The Ising model}\label{subsec:Ising}

The studies above demonstrate that the $2$d orders are a great testing ground to explore causal sets using computer simulations.
This can be taken even further.
One big question in causal set theory is, how to incorporate matter in the theory, and the $2$d orders let us explore this.
In particular, we can define an Ising type model, by assigning a `spin', a variable with values $[+1,-1]$ to each causal set element.
The interaction term for these can then be defined using the link matrix $L_{ij}$
\begin{align}\label{eq:isingaction}
  S_{\text{Ising}}&=j \sum_{k,l \in \mathcal{C}} s_k L_{kl} s_l =j \sum_{k \prec l} s_k s_l \;, 
\end{align}
to obtain a spin interaction along links~\cite{Glaser_2018,Glaser_2021}.

This makes it possible to study how the systems of matter and geometry interact, whether they give rise to new phases, and how the transitions between these phases behave.
It is particularly salient to examine if the transitions might become of higher order, since smooth phase transitions allow for a continuum limit.
While a continuum limit is not necessary for causal set theory, it does give more analytical tools to examine a theory.

A priori the Ising model has three possible states, spins can be correlated along links, anti-correlated along links, or uncorrelated, while the $2$d orders have two phases we know so far, the random $2$d orders and the crystalline orders.
If these phases arise in all combinations, without giving rise to any new behaviors, we would expect to find $6$ phases in the coupled system

\begin{figure}
  \centering
  \includegraphics[width=0.6\textwidth]{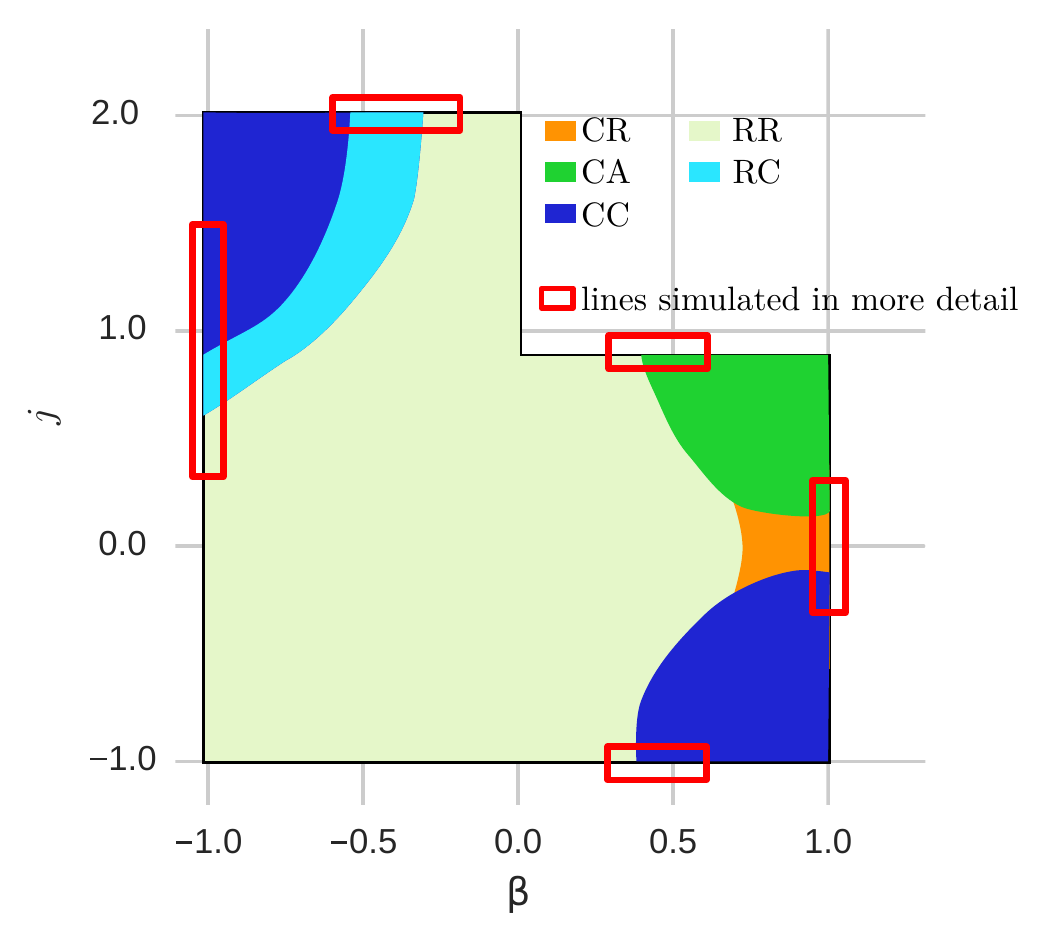}
  \caption{Phase diagram of the system of the Ising model coupled to the $2$d orders. The light green region RR are random $2$d orders with uncorrelated Ising spins, the vibrant green region CA are crystalline $2$d orders with anti-correlated Ising spins, the orange region CR are crystalline causal sets with uncorrelated Ising spins, and the dark blue region CC are crystalline causal sets with correlated Ising spins, the cyan region RC are random $2$d orders with correlated Ising spins. The red boxes mark lines that were studied in more detail. This figure is taken from~\cite{Glaser_2018}. }
\label{fig:isinPDiagram}\textbf{}
\end{figure}

To study these phases, in addition to the observables already used to study the $2$d orders, we also examined the Ising action~\eqref{eq:isingaction}, and the absolute magnetization of the system
\begin{align}
  | M | = |\frac{1}{N} \sum_i s_i | \;.
\end{align}
The phase structure as found in~\cite{Glaser_2018} is shown in Figure~\ref{fig:isinPDiagram}, where we see that only five different phases can be identified.
For the studies we explored $\epsilon=0.21$, since the prior studies of the $2$d orders without matter make us confident that there will be no qualitative change in the orders with $\epsilon$, as long as it is not chosen very small.

In the region around $\beta=j=0$ we find random $2$d orders with uncorrelated Ising spins, this is the region  where both systems are dominated by the entropy of configurations.
As $\beta$ increases while $j$ stays close to $0$ we find a transition to the crystalline $2$d orders, with uncorrelated Ising spins on them.
If $j$ increases we find crystalline causal sets with anti-correlated spins, while negative $j$ induces crystalline $2$d orders with correlated spins.

Since $\beta$ is a free parameter in this model we also extended the study to negative $\beta$.
We find that for $j=0$ and $\beta \in [0,-1]$ the $2$d orders still behave as random orders, without a clear phase transition.
The negative $j$ and negative $\beta$ region does not show any signs of a new phase arising.
However, this region should give rise to anti-correlated spins, which are harder to examine using the observables defined above, so it is possible that interesting behavior has not been found.

For positive $j$ and negative $\beta$ we find the most intriguing behavior.
Without the Ising model the path sum in this region would still be dominated by random orders, while the Ising action favors maximizing the number of links, and thus creating crystalline orders.
These competing effects give rise to a double phase transition from random $2$d orders with disordered spins, to random $2$d orders with ordered spins, to crystalline $2$d orders with ordered spins, which is a matter induced phase transition of geometry.

This process and the order of this induced phase transition was studied closer in~\cite{Glaser_2021}.
To do so we simulated the system at different sizes ranging from $N=20$ to $N=120$, for a line at $j=-1$ and one line along $j=1$.

Along the line~$j=-1, \beta \in [0,0.8]$ we find one phase transition, from a phase in which both the Ising spins and the geometry are random, into a phase in which the Ising spins are aligned, and the causal set is crystalline.

Studying this phase transition in detail we found that the phase transition happens at smaller $\beta$ values than for the pure $2$d order system, and scales like
\begin{align}
\beta_c(N)= (3.35 \pm 0.15 )\cdot N^{-0.72 \pm  0.01} \;.
\end{align}
This odd scaling in $N$ is likely an effect of the Ising model slowing the scaling as compared to that expected for the pure $2$d orders.

\begin{figure}
  \subfloat[$B_{M}$]{
  \includegraphics[width=0.49\textwidth]{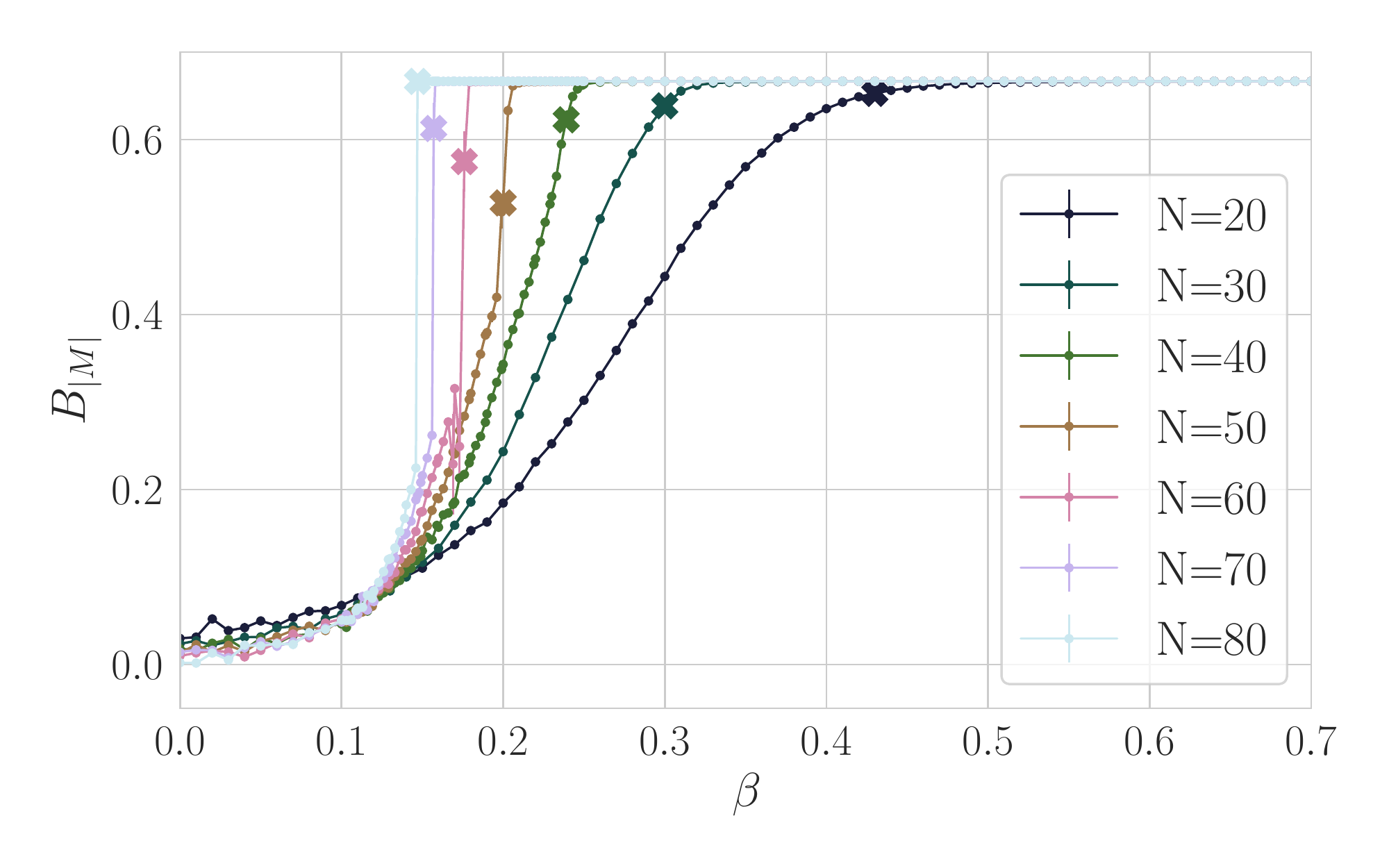}}
  \subfloat[$B_{S}$]{
  \includegraphics[width=0.49\textwidth]{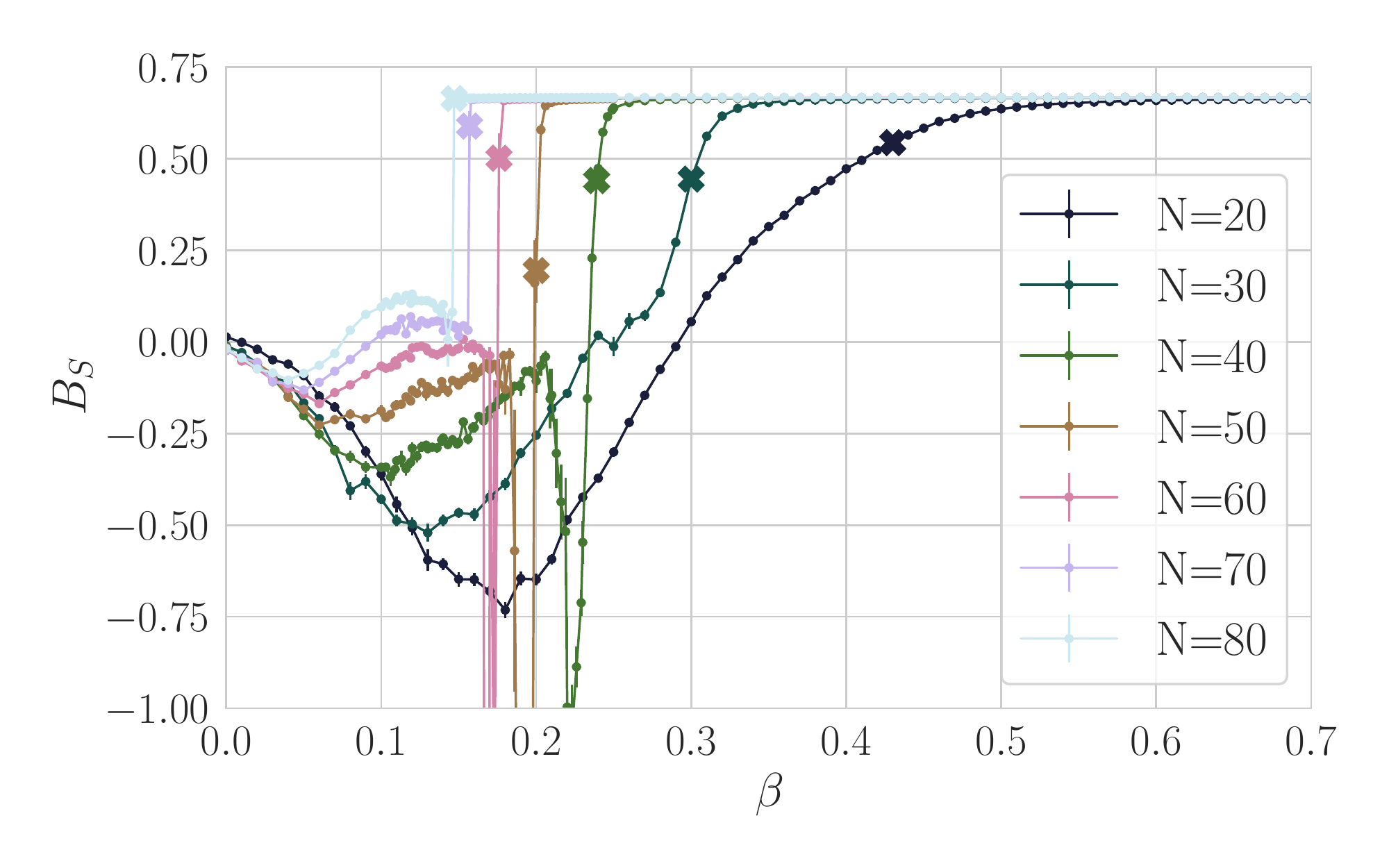}}

  \subfloat[$B_{S_{I}}$]{
  \includegraphics[width=0.49\textwidth]{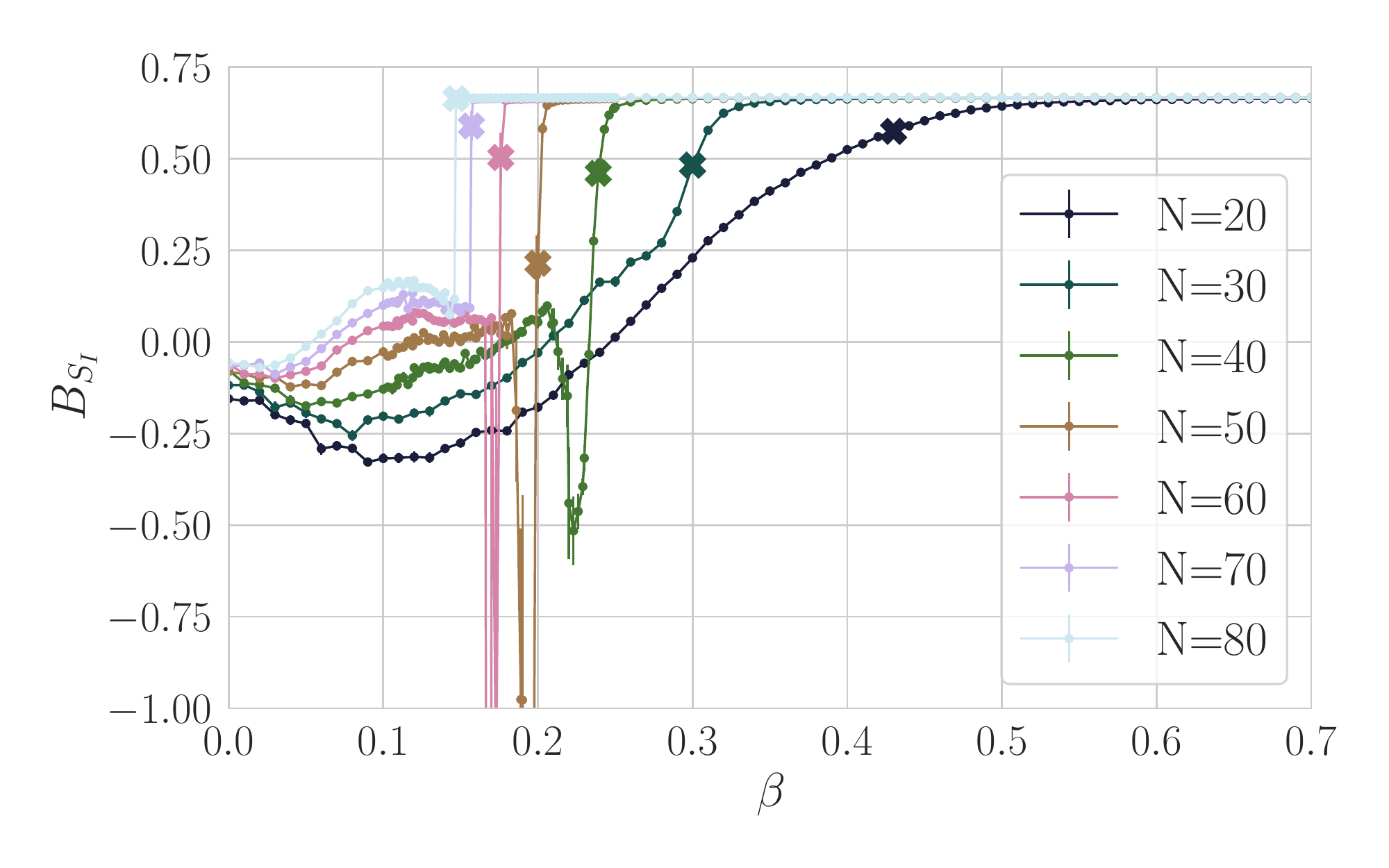}}
  \subfloat[$B_{S_{c}}$]{
  \includegraphics[width=0.49\textwidth]{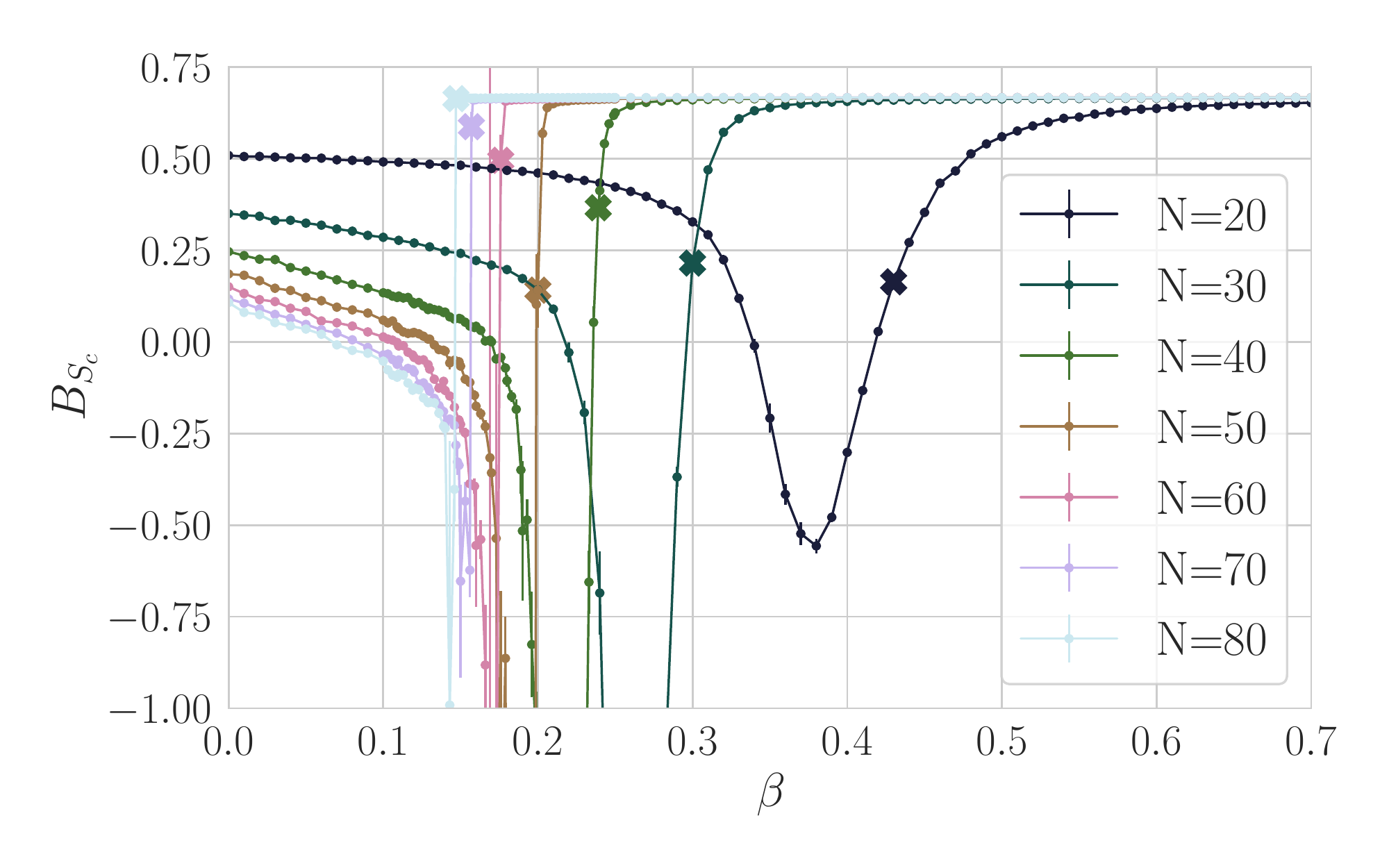}}

  \caption{The fourth order cumulant for the actions and the magnetization at $j=-1$. The thicker crosses indicate the location of the phase transition, as determined from the peak of $\mathrm{Var}(S)$.}\label{fig:Isingtransitionj-1}
\end{figure}

Looking at the Binder cumulant of the observables in Figure~\ref{fig:Isingtransitionj-1} we see that the transition appears of first order in the observables associated with geometry, but of higher order in those associated with the Ising spins.

This leads us to the conclusion that this is a mixed order phase transition.
While the transitions remain different in the two parts of the system, one can not deny that the systems are strongly influencing each other, since the location of the transition, and the scaling of this location with the system size, are changed considerably compared to the uncoupled systems.

Studying the scaling of the system with size shows that the actions scale linearly in $N$  in the low $\beta$ region where causal sets and spins are random and quadratic in $N$ in the crystalline and correlated region at high $\beta$.

The other line explored is~$j=1, \beta\in [-1.4,0]$, which crosses two phase transitions, including one induced by the Ising model.
At high $\beta$ we observed a magnetic transition from a phase of uncorrelated spins and random $2$d orders to a phase of random $2$d orders and correlated spins, which is followed by a geometric transition as $\beta$ is further decreased.
This transition goes from random $2$d orders with correlated spins to crystalline $2$d orders with correlated spins.

The scaling of the transitions with $N$ can be fit as
\begin{align}
  \beta_{c,mag} &= (-1.22 \pm 0.20) \cdot N^{-0.41 \pm 0.04}\\
  \beta_{c,geo}&=  (-8.58 \pm 0.34) \cdot N^{-0.77 \pm 0.01}
\end{align}
Extrapolating these fits to very high $N$ the lines would meet somewhere between $N \sim 500 - 1000$, depending on the error bars of the fits.
However physically this seems to be unlikely, since the spins need to align before they can force the $2$d order into a crystalline structure, we expect that the extrapolation is flawed and that the distinct phases will persist at large $N$.

\begin{figure}
  \begin{minipage}{0.14\textwidth}
    (a) $B_{M}$
  \end{minipage}
  \begin{minipage}{0.85\textwidth}
  \includegraphics[width=\textwidth]{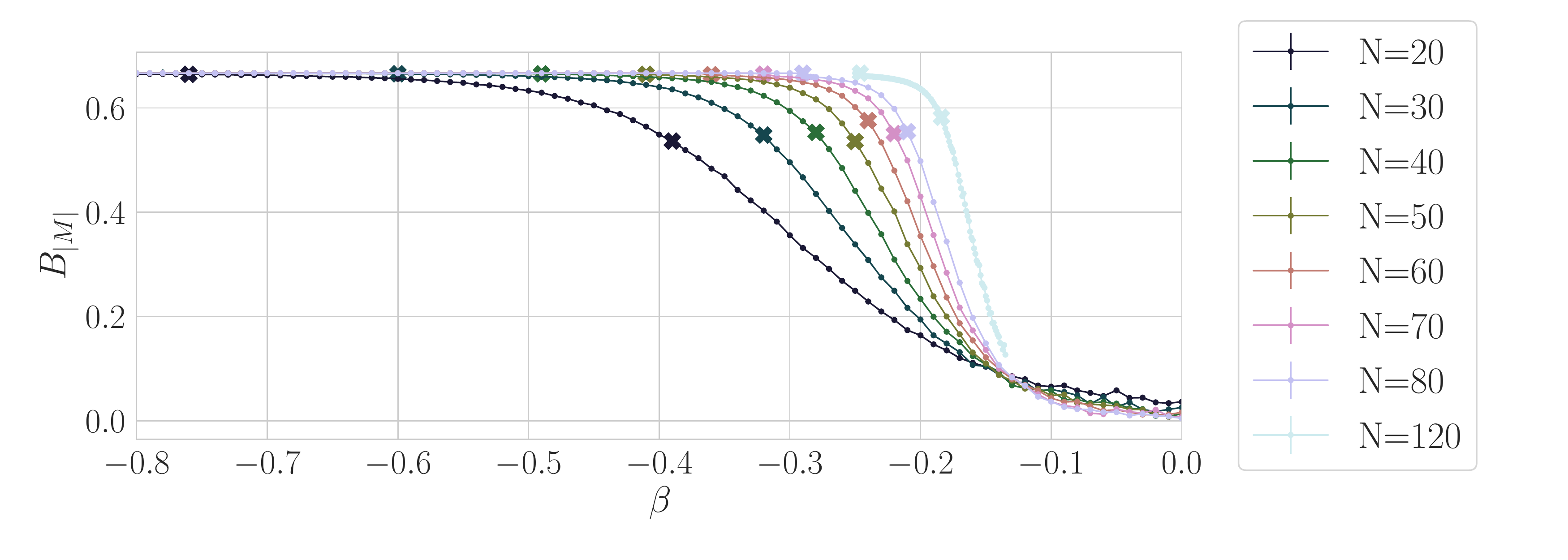}
  \end{minipage}
  \begin{minipage}{0.14\textwidth}
    (b) $B_{S}$
  \end{minipage}
  \begin{minipage}{0.85\textwidth}
    \includegraphics[width=\textwidth]{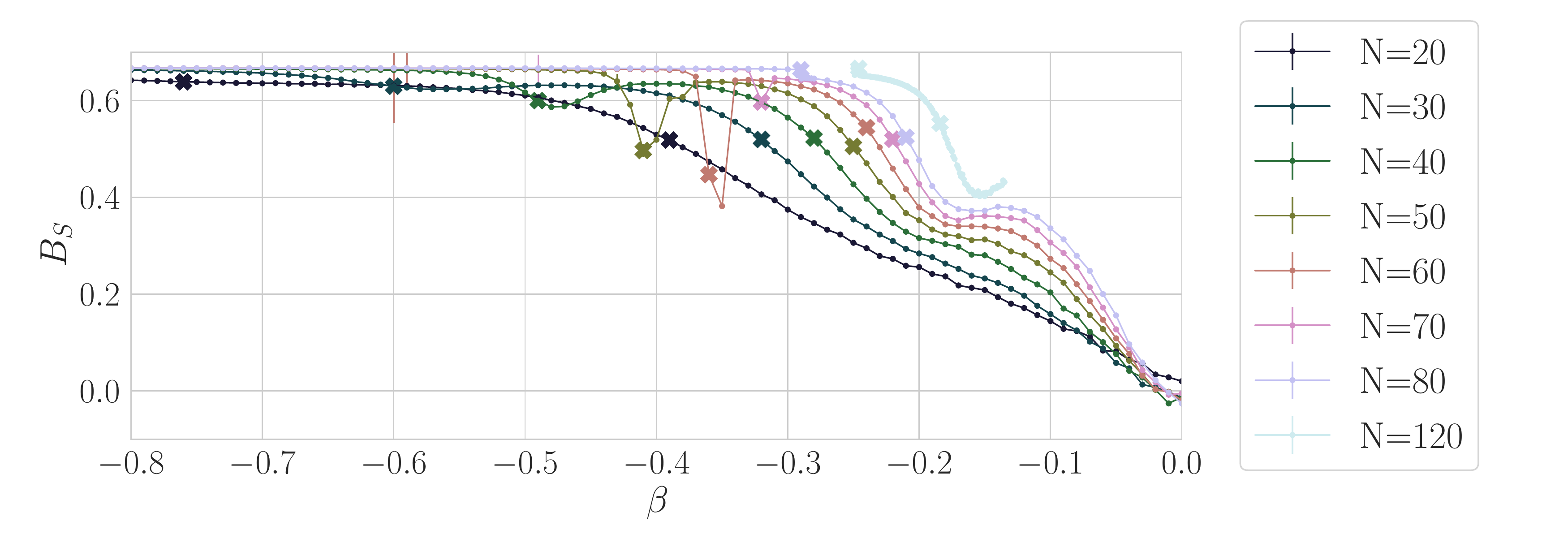}
  \end{minipage}
  \begin{minipage}{0.14\textwidth}
    (c) $B_{S_{I}}$
  \end{minipage}
  \begin{minipage}{0.85\textwidth}
    \includegraphics[width=\textwidth]{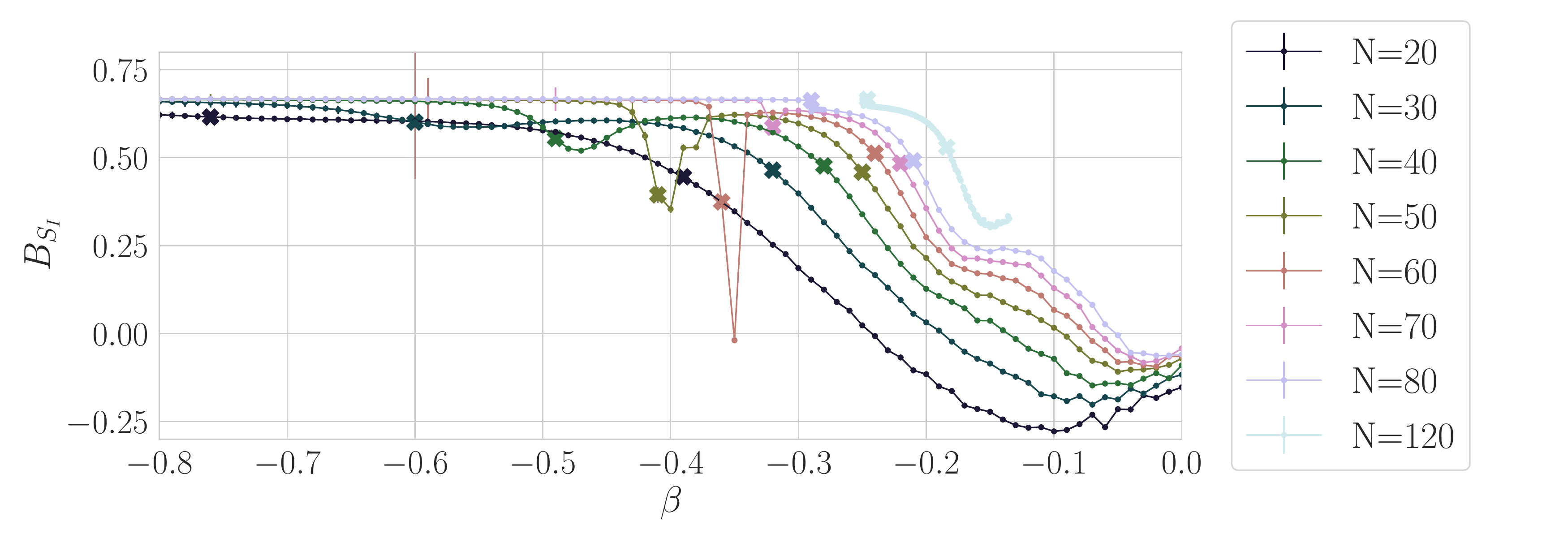}
  \end{minipage}
  \begin{minipage}{0.14\textwidth}
    (d) $B_{S_{c}}$
  \end{minipage}
  \begin{minipage}{0.85\textwidth}
  \includegraphics[width=\textwidth]{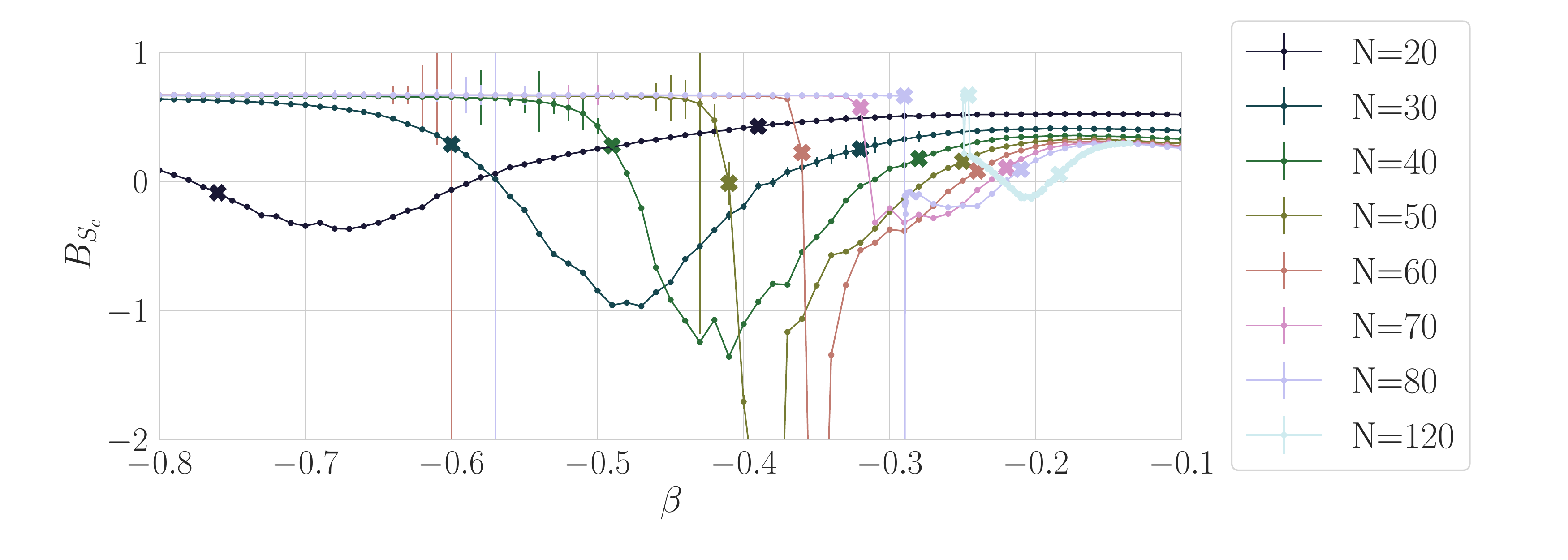}
  \end{minipage}

     \caption{Fourth order cumulant plotted against $\beta$ for both the action and the magnetization for $j=1$. The thicker crosses indicate the two phase transition points. The rightmost cross of each color is the magnetic transition, while the second cross indicates the geometric transition, both measured using the peaks in the Variance $\mathrm{Var}(S)$.}\label{fig:binder_j1_combine}
  \end{figure}

Exploring the transitions closer, we see that the observables are consistent with a higher order phase transition, apart from the histogram of the action whose distribution grows wider as $N$ increases, instead of peaking sharply.
The transition of geometry however still is appears consistent with a first order transition.
Both phase transitions can be seen in the Binder coefficients shown in Figure~\ref{fig:binder_j1_combine}.
So while matter can induce a phase transition in geometry, it does not fundamentally change the type of transition that geometry undergoes, in this context.

Examining the scaling of the action it is found to be identical to that of the random uncorrelated and the crystalline correlated phases respectively along the negative $j$ line.
For the region of random causal sets with correlated spins we did not find any clear scaling.

The work thus shows that coupling the Ising model to the $2$d orders does not fundamentally disrupt the scaling behavior.
While we do find some new phase transitions, they do not seem to be of higher order in geometry, and do thus not resolve the problem of predictability mentioned above.
It is however promising that they do show strong coupling between matter and geometry, thus allowing us to study a coupled system.
More work is certainly needed to understand more realistic matter on more realistic causal sets.

\section{Outlook}\label{sec:pers}
This chapter has introduced MCMC simulations on the $2$d orders as a model system of causal sets.
Even this simple case still leaves many interesting open questions for future investigation.
First we should further our understanding of matter on this simple system.
This could take the form of including a scalar field, either using the d'Alembertian operator defined above~\eqref{eq:boxricci}, or using the Greens function as described in the chapters on quantum field theory on causal sets by Nomaan X and Ian Jubb.
It would also be physically interesting to see how the wave function of the universe is influenced by the inclusion of matter.

The model of the $2$d orders can also be extended, allowing for more general configurations and thus moving closer towards the physically interesting case of $4$ dimensional spacetime.
This is done in the lattice gas model defined in~\cite{Cunningham_Surya_2020}.
One can think about a $N$ element $2$d orders as a lattice, with $N^2$ lattice sites of which only $N$ are occupied, but each occupied site precludes other elements sharing the same light-cone coordinates, which since there are only $N$ possible values for each coordinate means that with $N$ elements there can be no additional elements added still satisfying the conditions.
The generalized  lattice gas extends the configuration space by allowing for elements to share coordinates, and by decoupling the number of lattice sites from the number of elements.

This also allows for an extension to causal sets in higher dimensions.
In general extending the $2$d orders to $3$ or higher dimensions is not a good model of spacetime, since it would lead to `hypercube light cones'.
This implies that causal sets that approximate $3$d spacetime are not included in the $3$d orders, since their causality structure is not represented.
However, the more general lattice models allow for empty lattice sites, and thus do include manifoldlike configurations.
In particular the limit of taking the number of lattice sites to infinity for a fixed size of the causal set recovers the case of a sprinkling into flat Minkowski space.
The generalized lattice model also opens the possibility of exploring changes in the topology of our configurations.

\printbibliography

\end{document}